\documentclass[journal]{IEEEtran}
\usepackage{cite}

\ifCLASSINFOpdf
\usepackage[pdftex]{graphicx}
\else
\usepackage[dvips]{graphicx}
\graphicspath{{../eps/}}
\DeclareGraphicsExtensions{.eps}
\fi
\usepackage{epstopdf}
\usepackage{amsmath}
\usepackage{amsfonts,amssymb}
\usepackage{times}
\usepackage{tabularx}
\usepackage{booktabs}
\usepackage{algorithm}
\usepackage{algorithmic}
\usepackage[algo2e]{algorithm2e}
\usepackage{subfigure}
\usepackage{multirow}
\usepackage{url}
\usepackage{changes}
	\definechangesauthor[name={Yuanhao Cui}, color=orange]{per}

\begin{document}

%opening
\title{Decentralized Blockchain for Privacy-Preserving Large-Scale Contact Tracing}
\author{Wenzhe~Lv, \textit{Student Member, IEEE}, Sheng~Wu, \textit{Member, IEEE}, Chunxiao~Jiang, \textit{Senior Member, IEEE},  \\Yuanhao~Cui, \textit{Member, IEEE}, Xuesong~Qiu, Yan~Zhang, \textit{Fellow, IEEE}
\thanks{Wenzhe~Lv, Sheng~Wu, Yuanhao~Cui, Xuesong~Qiu are with School of Information and Communication Engineering, Beijing University of Posts and Telecommunications, Beijing 100876, China. (E-mail: \{lvwz, thuraya, cuiyuanhao, xsqiu\}@bupt.edu.cn)% \thanks{M. Shell was with the Georgia Institute of T echnology.}
	
Chunxiao~Jiang are with Space Center, Tsinghua University, 
Beijing 100084, China.  (E-mail: 
jchx@tsinghua.edu.cn)

Yan~Zhang are with Department of Informatics, University of Oslo, 
Oslo, Norway.  (E-mail: 
yanzhang@ieee.org)
}}
%Formatter's Note: Please include all the author names in the \author{} field above.

%\IEEEcompsocitemizethanks{\IEEEcompsocthanksitem }
%Formatter's Note: Please include the affiliation of each author, including the department, institution, city, state (if applicable), country, postal code, and email address in the \IEEEcompsocitemizethanks{} field above.
%Formatter's Note: Please include the corresponding author's name, mailing address, telephone and fax numbers, and email address here.

\markboth{Journal of \LaTeX\ Class Files,~Vol.~14, No.~8, August~2015}%
{Shell \MakeLowercase{\textit{et al.}}: Bare Demo of IEEEtran.cls for IEEE Transactions on Magnetics Journals}

\maketitle
\begin{abstract}
Activity-tracking applications and location-based services using short-range communication (SRC) techniques have been abruptly demanded in the COVID-19 pandemic, especially for automated contact tracing. The attention from both public and policy keeps raising on related practical problems, including \textit{1) how to protect data security and location privacy? 2) how to efficiently and dynamically deploy SRC Internet of Thing (IoT) witnesses to monitor large areas?} To answer these questions, in this paper, we propose a decentralized and permissionless blockchain protocol, named \textit{Bychain}. Specifically, 1) a privacy-preserving SRC protocol for activity-tracking and corresponding generalized block structure is developed, by connecting an interactive zero-knowledge proof protocol and the key escrow mechanism. As a result, connections between personal identity and the ownership of on-chain location information are decoupled. Meanwhile, the owner of the on-chain location data can still claim its ownership without revealing the private key to anyone else. 2) An artificial potential field-based incentive allocation mechanism is proposed to incentivize IoT witnesses to pursue the maximum monitoring coverage deployment. We implemented and evaluated the proposed blockchain protocol in the real-world using the Bluetooth 5.0. The storage, CPU utilization, power consumption, time delay, and security of each procedure and performance of activities are analyzed. The experiment and security analysis is shown to provide a real-world performance evaluation.
%Witnesses are rewarded from 
%including to provide activity-tracking services, to broadcast transactions, and to store location information. The transparent data storage
 %However, the computation cost of coverage optimization is not affordable for single IoT devices, especially when number of sensing devices is massive. Meanwhile, the transparent blockchain data raises more critical identity privacy issues. Therefore, we firstly propose a artificial potential fields based incentive mechanism to maximizing total coverage in a distributed manner. Secondly, we address the identity privacy issue by proposing a combination technique of a zero-knowledge proof and a key escrow. The resulting system is able to decouple the connection of the unique cryptographic identity and the on-chain location data such that it is almost impossible to track and identify the ownership of exposed on-chain location information. Furthermore, the owner of the location data is able to claim ownership without revealing the private key to anyone else. The security is also analysed to prevent various attacks. 
\end{abstract}

\begin{IEEEkeywords}
	Blockchain, Potential Field, Contact Tracking, COVID-19
\end{IEEEkeywords}

% To allow for easy dual compilation without having to reenter the
% abstract/keywords data, the \IEEEtitleabstractindextext text will
% not be used in maketitle, but will appear (i.e., to be "transported")
% here as \IEEEdisplaynontitleabstractindextext when the compsoc 
% or transmag modes are not selected <OR> if conference mode is selected 
% - because all conference papers position the abstract like regular
% papers do.
% \IEEEdisplaynontitleabstractindextext has no effect when using
% compsoc or transmag under a non-conference mode.

% For peer review papers, you can put extra information on the cover
% page as needed:
% \ifCLASSOPTIONpeerreview
% \begin{center} \bfseries EDICS Category: 3-BBND \end{center}
% \fi
%
% For peerreview papers, this IEEEtran command inserts a page break and
% creates the second title. It will be ignored for other modes.

%Formatter's Note: The journal limits this manuscript to 18 pages. Currently, your manuscript contains 29 pages. Please consider reducing the page count before submission to the journal.
\section{Introduction}
\subsection{COVID-19 Requirement}
The COVID-19 pandemic has been a global health emergency, spreading to more than 18 countries of the world and infecting more than 8,359,869 population. As an essential tool for public health officials and local communities to fight the rapidly spreading, contact tracing methods draw attention across the world. Early in the outbreak, when there were only a few cases, contact tracing can be done manually with a slight impact on society. Nevertheless, with the skyrocketing of the infections in the vast majority of countries,  manual contact tracing tends to be unrealizable on both economy and policy considerations. 

Location-Based Services (LBS) provided by Short-Range Communication (SRC) techniques are considered a viable solution to perform contact tracing. Due to the intensive penetration of mobile smartphones, wearable devices, and ubiquitous sensors, residents can opt-in contact tracing service via an over-the-air mobile application installation, whenever Internet access is available. The Singaporean government released a mobile phone app, TraceTogether, that is designed to assist health officials in tracking down exposures after an infected individual is identified. 

In general, a recorded SRC interaction with timestamp and locationinfo can certify that, at a given moment and position, two radio terminals have been closer than the maximum transmission distance. If one of the two terminal owners is tested positive for COVID-19, the SRC record can be strong evidence of epidemiological exposure for the other owner. Therefore, the contact identification can be automatically made by analyzing the infected person's SRC records. 

On the other hand, the chosen of SRC standard and storage methods would significantly impact the performance of SRC-based contract tracking.
% As the intensive penetration of mobile smartphones, wearable devices and ubiquitous sensors, out of the box SRC standards in these extensive electronics can be classified into three categories:
%\begin{itemize}
%	\item  Wireless Wide Area Networks (WWAN) such as Long Time Evolution (LTE), which are maintained by regional operators with (roughly speaking) up to 2km maximum transmission distance. 
%	\item Wireless Local Area Networks (WLAN) such as IEEE 802.11 standards,  commercially named Wi-Fi, maintained by individuals and on unlicensed spectrum, with up to 1.5km transmission distance. 
%	\item Wireless Personal Area Networks (WPAN) such as BLE, which can be maintained by telephones with about $10 \sim 20$m transmission distance and kept always-on with low power consumption. Take Aruba's BLE Beacon\cite{aurora} as example, its buttery life is up to 4 years. 
%\end{itemize}
% In general, the deployment of WWAN- and WLAN-type SRC techniques are relatively costlier in firmware customization and hardware update, which may not afford by deficit-ridden governments, especially in several third-class countries. On the other hand, 
Shorter communication distance increases identity accuracy. An overlong communication range may result in excessive identified contactees whose physical exposure may do not actually occur. As an example, the South Korea government analyzes Call Detail Record (CDR) collected from telecommunication operators. When a COVID-19 patient is detected, the entire building should be quarantined, which results in a significant impact on the company business. Therefore, benefit from easy to deployment and stone-cast data transmission, Bluetooth Low Energy (BLE) has been considered as a possible contact tracing approach.

\subsection{Daily Requirement}

If we take a longer-term and world-wide perspective, increasingly more people rely on LBS apps to track, monitor, manage, and plan their daily life. These mobile apps have profoundly impacted and even initialized various industries, such as Uber \cite{angrist2017uber} and DiDi in transportation sharing, Ele in food delivery, Fitbit and Nike+ in fitness, and Pokergo in augmented reality games.
%Editor: Please ensure that the intended meaning has been maintained in the edits in the previous sentence.
The key features of these apps are to collect location information in a certain time period and to perform path generation to share the personal location/activity path/summary with other users/service providers on a social network. For example, activity-tracking programs are developed in United Health and Aetna to modify insurance rates according to an activity analysis \cite{bertoni2014oscar}.

However, these activity-tracking services assume users are trusted and honest to report their location and time information, which is generally impractical and unrealistic. Indeed, the fashion of self-reporting location information using global positioning system (GPS) coordinates, cell triangulation and IP address tracking are all susceptible to manipulation such that users can handily claim a mendacious location \cite{doolittle2013ut}, which may cause serious cheating issues, especially to financial and critical service providers. To address these issues, SRC records also consider being an applicable method to improve their location reliability.

%Currently, a considerable amount of activity and SRC-based services are heavily tied to various financial incomes. The LBS apps try to balance supply and demand allocation to satisfy customer requirements of substantial professions, for example, to encourage far-distance drivers to accept ride orders in an area where there is a heavy flow of commuters such that drivers may be rewarded with financial interests \cite{xu2018large}.

 \subsection{Challenges and Motivations}

Although LBS provided by SRC (especially BLE) techniques are highly demanded in the contact tracing of COVID-19 pandemic and many more daily demands, much uncertainty still exists on data security, privacy, and deployment efficiency. 
\begin{itemize}
	\item Security: Traditional security services such as authentication, integrity, and provenance provided by third-party brokers. However, the third-party brokers maybe not trustable.
	\item Privacy: User privacy may be violated when service providers collect, store, and analyze customer's locations using a centralized database.  The service providers can even sell customer data by taking a little advantage of the fine print in the service agreements. Although Europe's General Data Protection Regulation (GDPR) has been implemented, the unintended data leakage of service providers still call for technical privacy concerns.
	\item Deployment Efficiency: A phone-to-phone SRC record demonstrates the spatial connection between two individuals. However, long-term monitoring of a high-risk hot-spot area is still critical in contact tracing. Governments or philanthropic organizations are willing to deploy several IoT witnesses, e.g., BLE beacons, to achieve an all-weather day-and-night low-cost area surveillance. These organizations may not cooperate. Therefore, an efficient deployment strategy is desired in the blockchain system.

	% Due to self-organizing and competitive transmission mechanism of Bluetooth, the majority of IoT devices are not able to perform centralized deployment algorithms. Therefore, how to efficiently deploy IoT witness in a distributed manner with coverage consideration has been an urgent problem.
\end{itemize}

%\replaced{These three major challenges motivate us to exploit a technical solution to alternate the trusted third-party-based protection and to build an activity-proofing, privacy protection system. Additionally, this system could be self-deployment. }{}

%需要再改
%A straightforward solution to these issues consists in enforcing the use of either secure or privacy-preserving location proofs for users [], [], [], [], where their location could be either (1) trusted and known (as it is the case for activity trackers) or (2) untrusted and known (but useless for obtaining rewards). In fact, solutions guaranteeing (1) would benefit the service provider by ensuring that cheating is infeasible, whereas solutions satisfying (2) would protect users’ location privacy but would provide locations that are too coarse-grained for computing meaningful summaries.

In this paper, we propose a permissionless blockchain for automated contact tracing and activity-proofing, named \textit{Bychain}. An interactive zero-knowledge proof based protocol is designed to provide transaction security and identity privacy while taking advantage of cryptographic techniques instead of employing a trusted third party. The deployment problem is addressed by proposing an incentive allocation mechanism based on a visual potential field algorithm; a participant would be distributed the maximum reward when it is on the position of force equilibrium. Ideally, the blockchain network archives maximized monitoring area when the IoT witnesses in the blockchain are static equilibrium.

This protocol relies on the cryptography technique to create SRC proofs for its nearby mobile users. The proposed protocol consists of three stages: First, users obtain secure and privacy-preserving proofs of location during their activities, by relying on an over-the-air lightweight message exchange between their mobile device and the witness point. Proof of Location (PoL) commitments are generated, digitally signed, and verified by their nearby witness nodes through BLE techniques. Second, the activity proofs are uploaded on a decentralized ledger that is transparently and distributively stored on the Internet. Finally, the trusted activity summary can be generated by using a zero-knowledge proof based authentication based on a smart contract after an ownership verification. The blockchain could compute an accurate and trusted activity summary of an authenticated user to fulfill the requirement of COVID-19 contact tracing, without revealing activity information to anyone else.

To empower Bychain with the ability of witness deployment, an incentive allocation algorithm is proposed based on the virtual potential field. Each witness node is treated as a charged particle, such that artificial electric fields are constructed in a way that each node is repelled by repulsive force from other nodes. As a result, the witnesses in blockchain could spread itself throughout the environment. Finally, the monitoring area of the blockchain system tends to maximize the monitoring area when the IoT witnesses in blockchain tend to static equilibrium.

Additionally, in the blockchain system, a semi-trusted decentralized activity-proofing system and zero-knowledge based authentication techniques could provide security from cheating and protect identity privacy. The public essential cryptography technique is naturally merged to provide an ownership certification of personal location information. As a benefit from the zero-knowledge proof technique, a user could claim data ownership without revealing the private key to anyone else. However, the decentralized ledger is transparent such that the public can verify the location information. Hence, we introduce a crucial escrow technique to alternate the traditional one-key-per-user with a one-key-per-proof scheme. The on-chain data traceability becomes almost impossible, and the identity information is pseudonymous. Thus, data security and identity privacy requirements are met in our system. Consequently, the attack of controlling the certificate authority and the central database can be avoided. Furthermore, plenty of attacks can be avoided, such as the reply attack and the dust attack. We will discuss them in detail later in this paper.
%\replaced{Benefit from blockchain system, in contrast to most of the existing schemes that require multiple trusted third parties to finalize authorization, storage and verification, our proposed framework does not rely on any trusted third party. Bychain is based on a decentralized ledger with communication that relies upon the cooperation among individual nodes to carry out essential tasks. }{}

%The block data of a traditional blockchain system are transparent and need to be verified by dynamic blockchain maintainers such that, ideally, anyone could access and analyze on-chain data. However, the data transparency violates user privacy protection because identity information is overt and can be tracked freely. To address this issue, we propose a key escrow technique and zero-knowledge proof-based on-chain verification technique to keep on-chain data anonymous but still variable without revealing the private key. The user's anonymity and location privacy is considered to be maximized in our framework.

The contributions of this paper can be summarized as follows:
\begin{itemize}
	\item A proof-of-activities generation and verification protocol is introduced to achieve trusted activities proof for contact tracing and related SRC usage. Bychain protocol addresses the activity-proofing problem without a trusted third party while considering privacy and anonymity. Buchanan can also fastly applied in COVID-19 with a low economic cost.
	\item By jointly design of crucial escrow and zero-knowledge proof method, the proposed protocol is able to obtain location identity privacy and security. It can resist various attacks and collusion. A generalized block structure is proposed to fulfill the requirement of new on-chain operations. Furthermore, to address the efficient deployment of IoT witnesses, an incentive allocation mechanism using virtual potential field is proposed to maximizing the monitoring area.
	\item A prototype implementation is realized and verified on the Android platform. The performance analysis shows that our protocol requires preferably low computational time, energy, and storage. Experiments show that the proposed protocol is practical and can be applied in many scenarios, even in a highly dynamic environment.
\end{itemize}

The rest of the paper is organized as follows. Section \ref{sec2} discusses related work. Section \ref{sec3} describes the overview of the proposed blockchain system. In Section \ref{sec4}, we discuss our proposed proof of activity protocol and incentive allocation mechanism. A security analysis of our proposed protocol against different types of attacks is provided in Section 5. In Section \ref{sec6}, we describe our implementation and simulation and present our experimental results on the performance evaluation. Finally, Section \ref{sec7} concludes the paper.

%Formatter's Note: Please provide Section VIII here.

\section{Related Work}\label{sec2}
Security and privacy of LBS and activity-tracking services are becoming a serious problem. The GPS verification mechanism can be modified by simply and easily cheating by modifying the geolocation API value return \cite{doolittle2013ut}. Moreover, the work in \cite{carbunar2012you} analyzed data from Foursquare and Gowalla and found that incentives to cheat exist because people actively check-in and collect rewards. Thus, it is necessary to carefully balance incentives with a more effective verification of users' location claims.

Many works have been discussed in the mobile computing society to offer secure verification of location information \cite{buttyan2007effectiveness,lenders2008location,pham2015securerun}. Short-range communication technologies, such as Bluetooth, have been proposed to generate location evidence from its neighbors. The mutual verification protocol of a set of users is proposed based on users' spatiotemporal correlation in \cite{talasila2010link}. A centralized location verifier is applied to verify mutually colocated users relying on Bluetooth communications. The privacy is performed by deciding whether to accept location proof requests and is decided by the users. APPLAUS is proposed in \cite{zhu2011toward} as similar work on a collusion-resistant location proof updating system using collocated Bluetooth devices. In addition, STAMP provides a spatial-temporal probabilistic provenance assurance for mobile users \cite{wang2016stamp}. The collision detection for such protocols is not 100 percent effective, and the protocol does not itself guarantee any collusion resistance. 

SRC based contact tracing is developing by Google and Apple \cite{googleapple} in a privacy-preserving fashion. However, the SRC records are stored in mobile terminals facing storage shortage problems. The deployment of IoT witnesses is not supported as well. Indeed, blockchain technology and its applications have drawn enormous attention from various researchers. In \cite{neural}, highly reliable communication is supported by blockchain and neural networks for unmanned aerial vehicles. A blockchain protocol is also proposed to manage data for the vehicular network in \cite{neural2}. However, the above applications can not address the location problems in this article.

%Physical layer communication information can be utilized to address the security and privacy issues.
%A location-aided routing protocol is proposed in \cite{li2010source}, where the current location of nodes is used to construct the network topology and forward data in an ad hoc networks. Multichannel state information from cellular telephony and satellite is used to determine the movement and location of user devices in \cite{courtney2000satellite}. Obfuscation-based techniques are discussed in \cite{duckham2005formal} to enable different degrees of location privacy based on varying the radius of a particular area.
\begin{figure*}[ht]	
	\centering
	\includegraphics[width=6.5in]{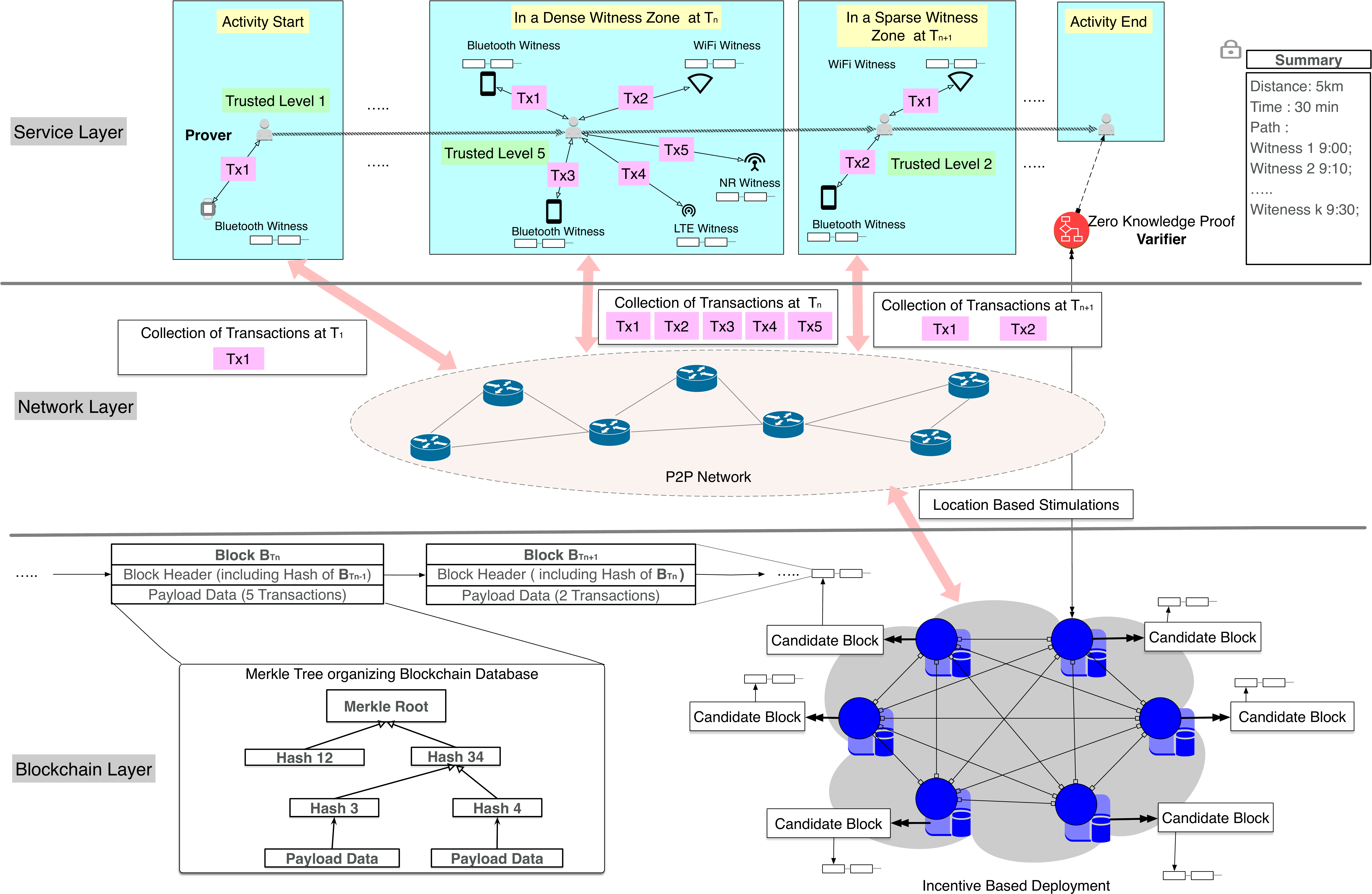}
	\caption{Overview of proposed blockchain architecture.}
	\label{fig:label}
\end{figure*}

\section{An Overview of Location-Based Blockchain Architecture}\label{sec3}
The proposed blockchain protocol is a peer-to-peer system, which provides users with anonymous and trusted location proofing, transparent data storage, and zero-knowledge proof based data ownership authorization. Fig. 1 illustrates the proposed blockchain architecture, in which the network is summarized as three layers : 
\begin{itemize}
	\item \textbf{The service layer} contains honestly witnesses that afford location-proofing service to passing-by provers via near-field communication techniques, and verifiers that finally generate activity certifications in the form of tabular activation summary. In a given time period, $T_n$, a prover collects witness-signed PoL commitments and then uploads them on blockchain as the cornerstone of trust evaluation. A straightforward trust level evaluation via the cumulative method is marked as green labels in Fig.1. 
	\item  \textbf{The network layer} enables the peer-to-peer (P2P) transmission mechanism to exchange collected PoL commitments of each prover and virtual force of each witness. This information are carried by the Transactions (Tx) field in the proposed generalized data structure and received by chain nodes. 
	\item  \textbf{The blockchain layer} is comprised of a distributed database recording immutable and continuously growing transactions and a location-based consensus mechanism maximizing total coverage of witness nodes, which is detailed in Sec.\ref{sec4}.  The blockchain system disburses a budget coverage package per 24 hours to stimulate witness movement to form maximum coverage of the sensing area. During each round of consensus procedure, a \textit{miner} calculates the incentive allocation scheme and then sticks it into the next block. The miner is pseudo-randomly determined from a node sequence in the order of owned stock value. The proposed block is broadcasted and finally confirmed by an elected consensus committee according to the value of mortgaged stocks. 
\end{itemize}

The key entities and mechanisms in each layer would be detailed in the following subsections.

\subsection{The Service Layer}

Individuals can access the blockchain with unified application software, which ensures data integrity by forcing a data consistency check when initializing. The service layer establishes a set of available application operations and coordinates the application's response in each operation. Although all IoT nodes with application installation are fully functional, the three logical roles involved in those operations need to be elaborated for a better understanding of readers.

\subsubsection{Prover}
In the proposed blockchain architecture, the prover acquires anonymous and trusted location proofing, transparent data storage, and zero-knowledge proof based data authorization. Provers indeed are a set of wireless nodes that desires location-based proof-of-activities. Each prover is equipped with a GPS module as well as Bluetooth, WiFi, or LTE antennas has Internet activity to the blockchain and moves in a given region that is covered by our wireless access point (AP)  networks. These provers can communicate with AP nodes involved in our system only if their distance is lower than the communication distance $R$, where $R$ is determined by the near-field communication technology, which means  Bluetooth,  WiFi or LTE will provide three kinds of distances, defined as $R_B, R_W$, and $R_L$, respectively. The sum of communication durations between provers and AP nodes can be completed before the prover leaves the coverage area of AP, which is verified in our experiment.

Each prover $i$ is equipped with a memory as a critical escrow agent to stock pairs of cryptology keys. The stocked vital pairs are generated when service initializes and randomly sign each near-field communication message, which means the $k$th proof-of-location commitment is uniquely identified by public key $pk_{p_i}^k$ and a private key $sk_{p_i}^k$. Without loss of generality, we assume provers will never hand or even reveal their private keys to anyone else.

\subsubsection{Witness}    
Witnesses are wireless nodes that provide location-based proofs to provers. These witnesses may be fixed WiFi APs, Bluetooth low power equipment (BLE) or an LTE base station owned by an ISP provider, deployed in the area where provers pursue their proofs. More importantly, a witness could discover another if their distance is lower than its coverage radius. Hence, witnesses are aware of the locations of their neighbors.

All the witnesses and provers have synchronized clocks and are equipped with a GPS device that is aware of the location of itself. This WiFi AP localization technique can be realized by the analysis of communication channel state information. Different from provers, each witness $i$ can be uniquely identified by a pair of cryptology keys, known as a public key $pk_{w_i}$ and a private key $sk_{w_i}$. Without loss of generality,  we also assume the witness will never hand or even reveal their private key or digital signature to anyone else. The witness can access the Internet and communicate with the blockchain.

In our system, we assume each witness is semi-trusted and privacy-honest, which means it does not reveal the prover's information to any nodes, but cheating such as collusion may occur. Practically, a witness is probably willing to collude with other witnesses or provers cheating for blockchain awards.

\subsubsection{Verifier}
A verifier is an entity that a prover wants to request an activity certification from, such that a prover's claim appears in a certain location at a particular time. The verifier is completed by interactive zero-knowledge proof based on the smart contract, such that it is self-running and able to protect the identity privacy of a specific user.

When the prover intends to request certification, a verification request including the on-chain index of PoL commitment, public keys of the prover, and witness are submitted to the blockchain system. The verifier smart contract is triggered when the particular verification request appears in a new block. Consequently, the ownership of the PoL commitment is verified by a zero-knowledge proof algorithm without revealing private keys to anyone else. Hence, the prover can be verified and certified as the owner of the particular PoL commitment, i.e., the prover is witnessed at a certain timestamp.

\subsection{The Network Layer}
The network layer takes responsibility for routing data packages between blockchain nodes. Considering lacking retransmission and reordering in User Datagram Protocol (UDP), transmission reliability of large package (block size up to 4 MB) could not meet the network requirement. Bychain uses an unstructured peer-to-peer network with TCP connections as its foundational communication protocol. 

\begin{figure*}[ht]    
	\centering
	\includegraphics[width=6.5in]{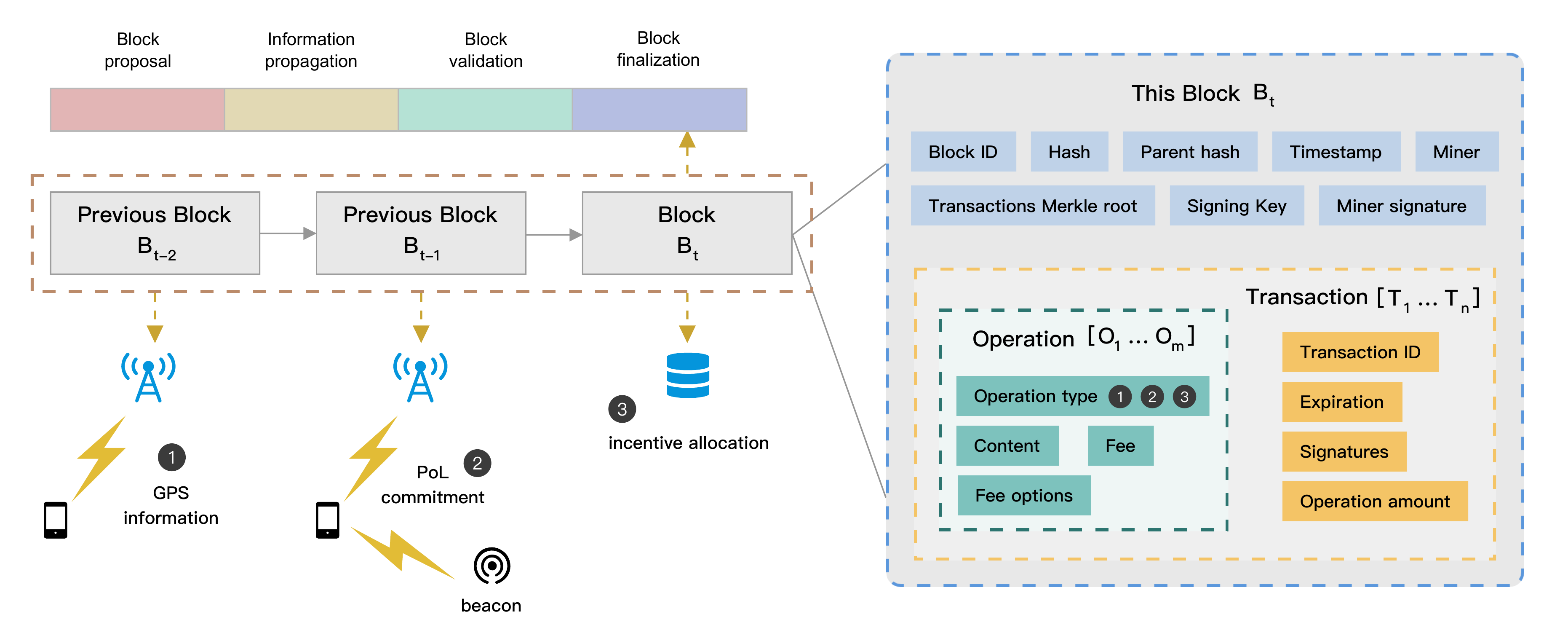}
	\caption{The proposed generalized block structure, the transaction field is extended to support multiple new operations.}
	\label{fig:2}
\end{figure*}

Bychain introduces several novel operations in the service layer such that traditional blockchain data structure \cite{zhou2018beekeeper} could not be straightforwardly applied. Take several new operations as an example, 1)  the prover intermittently broadcasts a package including multiple PoL commitments collected in a time window. 2) The witness reports its location and virtual forces during every round of consensus procedure. 3) In the blockchain layer, the miner publishes the \textit{incentive allocation} into the next block, which will be detailed in the next section.

Hence, for efficient and confidential data transmission, Bychain reconstructs the transaction structure and then develops a general block format, as shown in Fig.2. More precisely, \textbf{Expiration} and \textbf{Signature} fields are applied to indicate the producer and timeslot of one transaction. Hence, there would be multiple \textbf{Operations} happening over the specified identity and time duration. The broadcasted transactions are stored in a buffer pool to be serially packaged into \textbf{Blocks}. The buffer pool is maintained and checked by the consensus committee.

   %还有blockchain层的东西，包括最大化覆盖+Dpos的共识机制，秘钥管理和零知识证明的扩展

  \subsection{The Blockchain Layer}
  
  In contrast to existing location-proof methods, the proposed system does not require a trusted centralized third-party. Furthermore, aiming to endow the blockchain with maximizing the monitoring area, Bychain proposes a new location consensus based on the virtual electric field to maximizing the sum coverage of the location-proofing service. Furthermore, Bychain adopts an interactive zero-knowledge proof method to ensure data confidentiality and anonymity. We will describe the necessary entities and mechanisms in the following.
  
  Blockchain is composed of a distributed database and a peer-to-peer node network. The blockchain database is a secured, shared, fault-tolerant, distributed, and append-only database that facilitates consensus-based recording and tracking information without a centralized, trusted third party. The blockchain network is based on the peer-to-peer communication protocol and untrusted nodes.
  
  All the data processes on the blockchain are decided by the majority users in the blockchain network, and the decision-making procedure is called \textit{consensus generation} or \textit{mining a block}. All the network nodes attend a mining competition in each turn to decide which node can produce and broadcast the new block to the other nodes. The new blocks are composed of new unpacked data in the distributed database. There are several existing \textit{consensus generation} algorithms, including Proof-of-Work (PoW) \cite{nakamoto2008bitcoin}, Proof-of-Stake (PoS) \cite{saleh2018blockchain,kiayias2017ouroboros}, Proof-of-Space (PoSpace) \cite{nguyen2018survey} and practical Byzantine fault tolerance (PBFT) \cite{cachin2016architecture}.

  The blockchain system produces cryptocurrency to incentivize miners to take part in competitions for block recording rights in each turn. Due to the majority decision and resource-consuming competition, the blockchain does not rely on a traditional trusted third party when recording data if the majority of network nodes are honest. It is because a successful attack can only occur when the attacker wins enough mining competitions; however, the cost of this \textit{51\% attack} would go beyond expectation. According to an analysis in 2018 December \cite{bitcoinattack}, it will cost 1.4 billion dollars to realize the \textit{51\% attack} in the Bitcoin blockchain.
  
  Although a blockchain is composed of a database and a node network, the decentralized system is able to process some proper, on-chain, and heavily automated workflows by using the concept of the \textit{smart contract} \cite{buterin2014next}. \textit{Smart contracts} are self-executing scripts stored in the blockchain database and independently executed on each network node in a sandboxed virtual machine. The Turing-complete virtual machine allows us to implement complicated logic smart contracts with deterministic outputs and on-chain data interactions. The smart contract can be regarded as a predetermined on-chain rule triggered by particular on-chain data, such as a specific transaction.

  \subsubsection{Signature and Key Escrow}
  To validate the authenticity of PoL commitments and on-chain transactions, our blockchain employs the elliptic curve digital signature algorithm (ECDSA) asymmetric cryptography technique. The ECDSA cryptography is implemented by using the secp256k1-based Koblitz curve for the ECDSA key-pair in the open-source project OpenSSL. The generated private witness key is able to sign the PoL messages to authorize that the prover appears in a particular location and at particular timestamps. Furthermore, the on-chain transaction data also need to be signed such that the transaction is traceable and verifiable.
  
  Due to the on-chain data transparency, anyone can access and analyze the on-chain data freely. In this case, identity privacy becomes a critical problem introduced from the blockchain technique because the prover's activity information can be easily tracked on the blockchain.  In our protocol, a key escrow is employed for the prover to achieve the separation of identity information and location information. A primary private key and several generated key pairs are initialized and stored in the memory when the system initializes. The generated vital pairs are iteratively employed for each PoL message interaction procedure, and consequently, the generated public key is uploaded into the blockchain database for verification.  Because it is almost impossible to calculate the main private key based on the generated public key, we consider that the identity privacy is achieved.
  
  \subsubsection{Zero-Knowledge Proof Based Authorization}
  Because the key escrow is applied in our protocol, the only way to verify the identity information of the on-chain PoL commitment is by checking the private key. However, there is no third party that can be trusted to verify the private key on a blockchain system. Otherwise, location privacy may be 'stolen' by copying the private key.
  
  \textit{zero-knowledge proof} technique tries to help a verifier trust a prover without leaking any secret information to anyone else during the verification procedure. In this paper, we employ an interactive proof scheme in which the prover demonstrates its authorization information (private key) to the verifier by several interactive rounds. During the interactive procedure, the prover answers a randomly generated challenge from the verifier. However, most of the recent zero-knowledge proof based authorization to reply to a trusted third party, which is not always practical in the real world. 
  
  In our system, zero-knowledge proof based authorization procedure is considered part of our blockchain so that the information leakage risky from a third party is avoided. Hence, the private key is not required when verifying the clear on-chain proof-of-location commitments. 
  
  Furthermore, the blockchain itself can generate and distribute and update sets of pseudonyms (the pairs of the public key and private key as mentioned above) for every prover and witness.

\section{Proposed Protocol and Location-Based Consensus}\label{sec4}

In contrast to existing location-proof methods, the proposed system does not require a trusted centralized third-party. Furthermore, Bychain proposes a new DPoS consensus based on the virtual electric field to maximize the sum coverage of the location-proofing service. We will describe the necessary entities and mechanisms in the following.

\subsection{Proposed Protocol}
Our goal is to establish an activity-tracking and location-proofing system that guarantees the authenticity of the user location and activity data with respect to cheating nodes, the underlying untrusted centralized supernode, and malicious nodes, and provide almost complete privacy protection with respect to on-chain nodes that are trying to track provers.
In this section, we detail our blockchain-based proof-of-location system. The system is shown in Fig.~\ref{fig:label}, and the scheme is shown in Fig.~\ref{fig:label4}. Two stages are involved in our proof-of-activities protocol, the proof-of-location stage, and the activity generation stage, and they are denoted as follows,

\begin{figure*}[ht]
	\centering
	\includegraphics[width=7.5in]{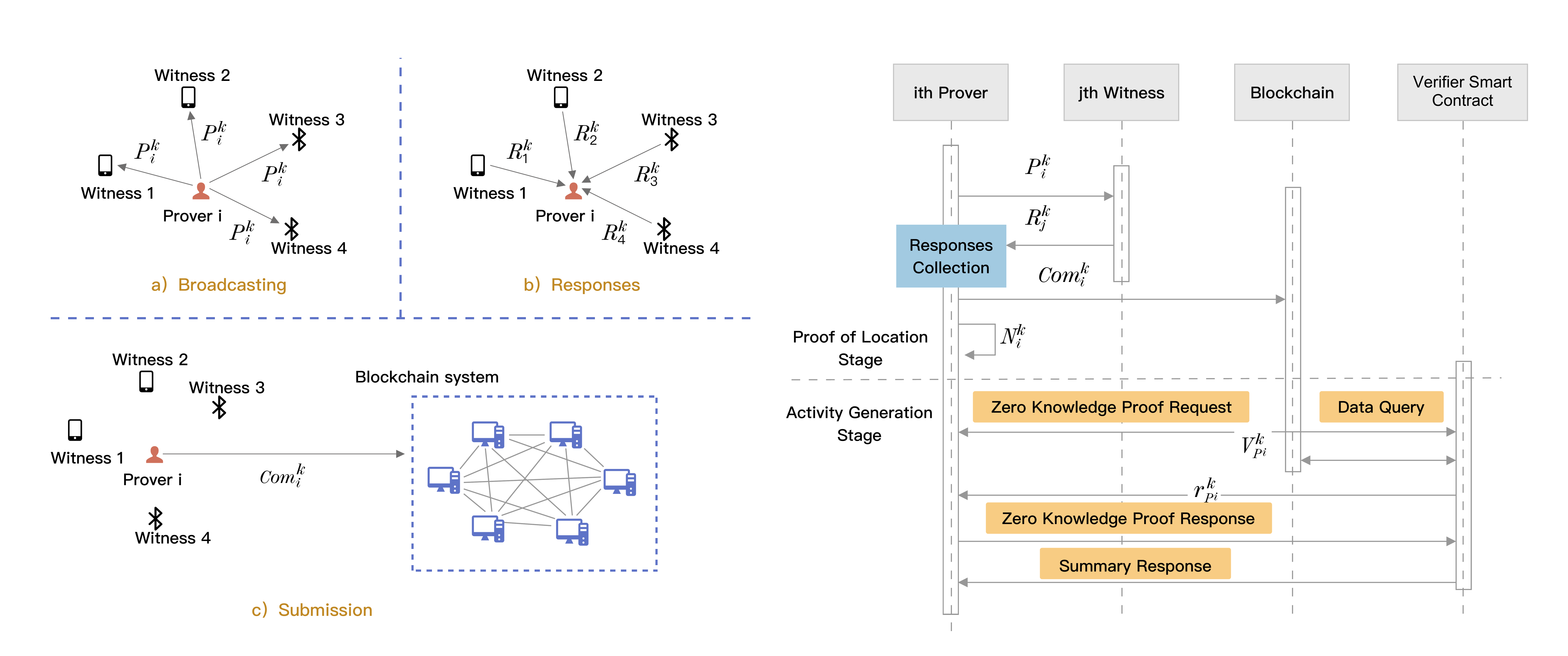}
	\caption{The proposed protocol consists of two stages, including PoL stage and activity generation stage.}
	\label{fig:label3}
\end{figure*}
~\\
\textbf{Stage 1: Proof of Location}
	
	\textit{1) Broadcasting}: The broadcasted PoL request can be descried by multiple witness nodes as shown in Fig.~\ref{fig:label3}.  Suppose a prover $p_i$ at location $L_{i,t}$ wants to start an activity proof collection event at sampling time $t$, so it periodically generated a location proof request $P^k_i$ using the $k$th pair of cryptology keys $\{pk_{p_i}^k,sk_{p_i}^k\}$ and broadcasts $P^k_i$ to nearby devices. Note that $\{pk_{p_i}^k,sk_{p_i}^k\}$ will be updated as $\{pk_{p_i}^{k+1},sk_{p_i}^{k+1}\}$ when this PoL stage is completed.  The PoL request is constructed as follows,
	\begin{equation}\label{requestgeneration}
	P^k_i = C\{pk_{p_i}^{k},\textrm{nonce}_{p_i}(t),L_{{p_i},t},t,S_{sk_{p_i}^k}\},
	\end{equation}
	where  $\textrm{nonce}_{p_i}(t)$ is a random number with respect to prover $p_i$ and timestamp $t$  and $L_{p_i,t}$ represents the current location information at time $t$. The payload data is compressed by the SHA256 function $\mathcal{H}()$ to reduce the communication complexity in the signature.  $S_{sk_{p_i}^k} = \textrm{Sign}_{sk_{p_i}}\{\mathcal{H}(pk_{p_i}^{k},\textrm{nonce}_{p_i}(t),L_{{p_i},t},t)\}$ denotes the signature of $sk_{p_i}^k$ agrees with the PoL request. The request is specified with respect to wireless protocols that are denoted as $C\{\}$.

	\textit{2) Response}: Each witness in its communication range decides whether to respond to the request. The witness node may reject the PoL request due to fake parameters, for example, the location information in $P_i^k$ shows that distance between the prover and current witness is significantly larger than the communication range. Request rejection also happens if crypto-verification fails, i.e.,  $S_{sk_{p_i}^k} \neq \textrm{Sign}_{sk_{p_i}}\{\mathcal{H}(pk_{p_i}^{k},\textrm{nonce}_{p_i}(t),L_{i,t},t)\}$ or is not paired with the public key $pk_{p_i^k}$ in $P^k_i$.  If the request is accepted by the witness $j$, the witness sends back a response message $R_j^k$ at timestamp $t'$. Without loss of generality, the time difference $t'-t$ is not enough for the prover to move out of witness $j$'s communication range. $R_j^k$ can be constructed as follows,
	\begin{equation}\label{1}
	R_j^k= C\{pk_{p_i}^{k},\textrm{nonce}_{p_i}(t),L_{i,t},t,S_{sk_{p_i}^k},\textrm{nonce}_{w_j},S_{sk_{p_i}^k,sk_{w_j}^k}\},
	\end{equation}
where $\textrm{nonce}_{w_j} = \{pk_j||n\}$ is a combination of public key $pk_j$ of witness $j$ and an unrepeatable built-in counter as $n$. It is worth noting that $pk_j$ is specific and  unique to $j$-user, and $\textrm{nonce}_{w_j}$ functions as an on-chain index of the PoL commitments. 

\begin{equation}\begin{aligned}
S_{sk_{p_i}^k,sk_{w_j}^k} = \textrm{Sign}_{sk_{w_j}}\{\mathcal{H}(pk_{p_i}^{k},\textrm{nonce}_{p_i}(t),L_{i,t},\\t,S_{sk_{p_i}^k},\textrm{nonce}_{w_j})\},
\end{aligned}\end{equation} 
denotes that a signature of $sk_{p_i}^k$ agrees with this response. This response is sent back to the prover as a witness location proof.
	
	%Finally, the prover construct a message $U_i$ and then, upload it on blockchain to make it immutable. The message $U_i$ is constructed as follows,
	%\begin{eqnarray}
	%U_i = C(|PK_i^{Pub},r_i(t),L_{i,t},t|_{WK_i^{Prv}}, |WK_j^{Pub},t',L_{j,t}|_{PK_i^{Prv}}),
	%\end{eqnarray}
	\textit{3) Submission}: Consequently, a dozen of responses with diverse witness signatures are collected at the prover's antenna in a short time window. If the responses originate from $M$ receivers, then the PoL commitment can be constructed by combining all the PoL responses,
	\begin{equation}\label{com}
	\textrm{Com}_i^k = \{R_1^k||R_2^k||\cdots||R_M^k\}.
	\end{equation}
	$\textrm{Com}_i^k$ is uploaded on the blockchain such that the PoL commitment for $k$th PoL request at the $i$th prover is immutable. A trusted level of $\textrm{Com}_i^k$ can be generated  at the activation summary depending on number of combined PoL responses. The prover-witness collusion attack can be avoided when $M$ is large enough and the honest witnesses are more than $\frac{M}{2}$. These steps are illustrated in Fig.~\ref{fig:label3}.

	For each submitting procedure $k$, a PoL note is stocked for a further zero-knowledge proof. The PoL note is a combination of private key $sk_{p_i}^k$ in the broadcasted request, the random number $\textrm{nonce}_{p_i}(t)$ and the collected witness indices $\textrm{nonce}_{w_1}, \ldots, \textrm{nonce}_{w_M}$ and can be written as,
	\begin{equation}\label{N}
	N_i^k = \{sk_{p_i}^k||\textrm{nonce}_{p_i}(t)||t||\textrm{nonce}_{w_1}||\cdots||\textrm{nonce}_{w_M}\}
	\end{equation}
	PoL notes the function as an identifier of its corresponding on-chain commitment, prover always can precisely locate the on-chain commitment and proofs of the ownership to anyone else by a zero-knowledge proof smart contract verifier without revealing the private key $sk_{p_i}^k$.

%Formatter's Note: Per the journal guidelines, all figures must be cited in the text. Currently, Figure 3 is not cited in the text. Please add a citation for this Figure or remove it from your manuscript and renumber the remaining Figures.

\begin{algorithm}[H]
	\caption{The Proposed Bychain Protocol}
	\LinesNumbered
	\KwIn{Key pair $\{pk_{p_i}^k,sk_{p_i}^k\}$, Responsed Witeness Set ${w_1,w_2,...,w_M}$, Maximum Response Duration $T_{max}$.}
	
	%\KwIn{$\textrm{Proof},S_{sk_{p_i}^k},\textrm{nonce}_{p_i}(t), L_{i,t},\textrm{nonce}_{p_i}(t),pk_i, t$}
	\KwOut{Activity Summary}
	\textbf{Stage 1 : Proof of Location}\ 
	
	\For{Prover $p_i$}{	
		Generate and Broadcast PoL request $P^k_i$ according to \eqref{requestgeneration}.
		A clock is initialized as $T$.
		
		\ForEach{witness $i$ in ${w_1,w_2,...,w_M}$}
		{PoL response $	R_j^k$ is generated according to \eqref{1}}
		
		\While{$T \leq T_{max}$}{Responses collection and combination according to \eqref{com}.}
		Local storage for private key and nonces storage according to \eqref{N}. 
	}
	\textbf{Stage 2 : Activity Generation}\ 
	
	\For{Prover $p_i$}{	
		Generate and transmit verification request according to \eqref{verify}.
		
		$r_{p_i}^k$ is returned from contract.
		
		Generate and transmit Proof according to \eqref{proof}.
		
		\For{Verifier}{
			\eIf{$\text{Verify}_{pk_i}(\text{Proof}) == \text{Sign}_{sk_{p_i}}\{\mathcal{H}(\text{nonce}_{p_i}(t), r_{p_i}^k)\}$}{
				%\State
				\eIf{$\text{Verify}_{pk_i}(S_{sk_{p_i}^k} )$ equals to $\text{Sign}_{sk_{p_i}}\{\mathcal{H}(pk_{p_i}^{k},\text{nonce}_{p_i}(t),L_{i,t},t)\}$}
				{Activity Summary transmission starts.}{The verification is failed.}
			}{The verification is failed.}
		}
	}
\end{algorithm}
~\\
\textbf{Stage 2: Activity Generation}
 	%A private key owner sends a request to verifier smart contract to obtain its activity summary. The verifier smart contract is avaliable on blockchain and can be access from any public API or self-build blockchain node. The contract search each block for personal data and then, combine each location proof into one table, by ordering them in time. This procedure is similar with remaining asset evaluation on blockchain. Then, the time, path can be calculated according to the table. Noting all these calculation is in the smartcontract such that it will not reveal to anyone else.
 	
 	In our system, the activity summary is generated by two steps: location verification and summary generation.

	\textit{1) Location Verification}: To verify the $k$th commitment, a single-round interaction zero-knowledge proof is applied. Firstly, prover $p_i$ produces a verification request,
	\begin{equation}\label{verify}
	V_{p_i}^k = \{ \textrm{nonce}_{w_1}||\cdots||\textrm{nonce}_{w_M}\}.
	\end{equation}   The smart contract verifier searches on-chain data according to index $\textrm{nonce}_{w_1}||\cdots||\textrm{nonce}_{w_M}$.  The outcome is the $k$th PoL broadcasting request, shown as $\{L_{i,t},\textrm{nonce}_{p_i}(t),pk_i, t, S_{sk_{p_i}^k}\}$. Consequently, the verifier generates a pseudorandom number $r_{p_i}^k$ as response to prepare zero-knowledge verification.
 		
 		Secondly, the prover signs a proof by using $sk_{p_i}^k$, as 
 		\begin{equation}\label{proof}
 		\textrm{Proof} = \textrm{Sign}_{sk_{p_i}}\{\mathcal{H}(\textrm{nonce}_{p_i}(t), r_{p_i}^k)\}.
 		\end{equation} The verifier can identify the ownership of the commitment by comparing $\mathcal{H}(\textrm{nonce}_{p_i}(t), r_{p_i}^k)$ with on-chain $\textrm{nonce}_{p_i}(t)$ and local $ r_{p_i}^k$, which is according to the randomness of $r_{p_i}^k$ and $\textrm{nonce}_{p_i}(t)$. The private key $sk_{p_i}^k$ will not be revealed to the verifier such that no one can complete the ownership verification in the future except prover $i$.		
 		
 		Finally, the verifier returns true or false as a result of PoL verification. It implies whether or not the PoL commitment belongs to the specific prover $i$. 
 		
	\textit{2)  Summary generation}: After each PoL is verified, the smart contract produces an activation summary, including the trusted level generated from the number of witnesses $M$ and the activation path.

\subsection{Location-based Consensus}

\begin{figure}[ht]		
	\centering
	\includegraphics[width=3.5in]{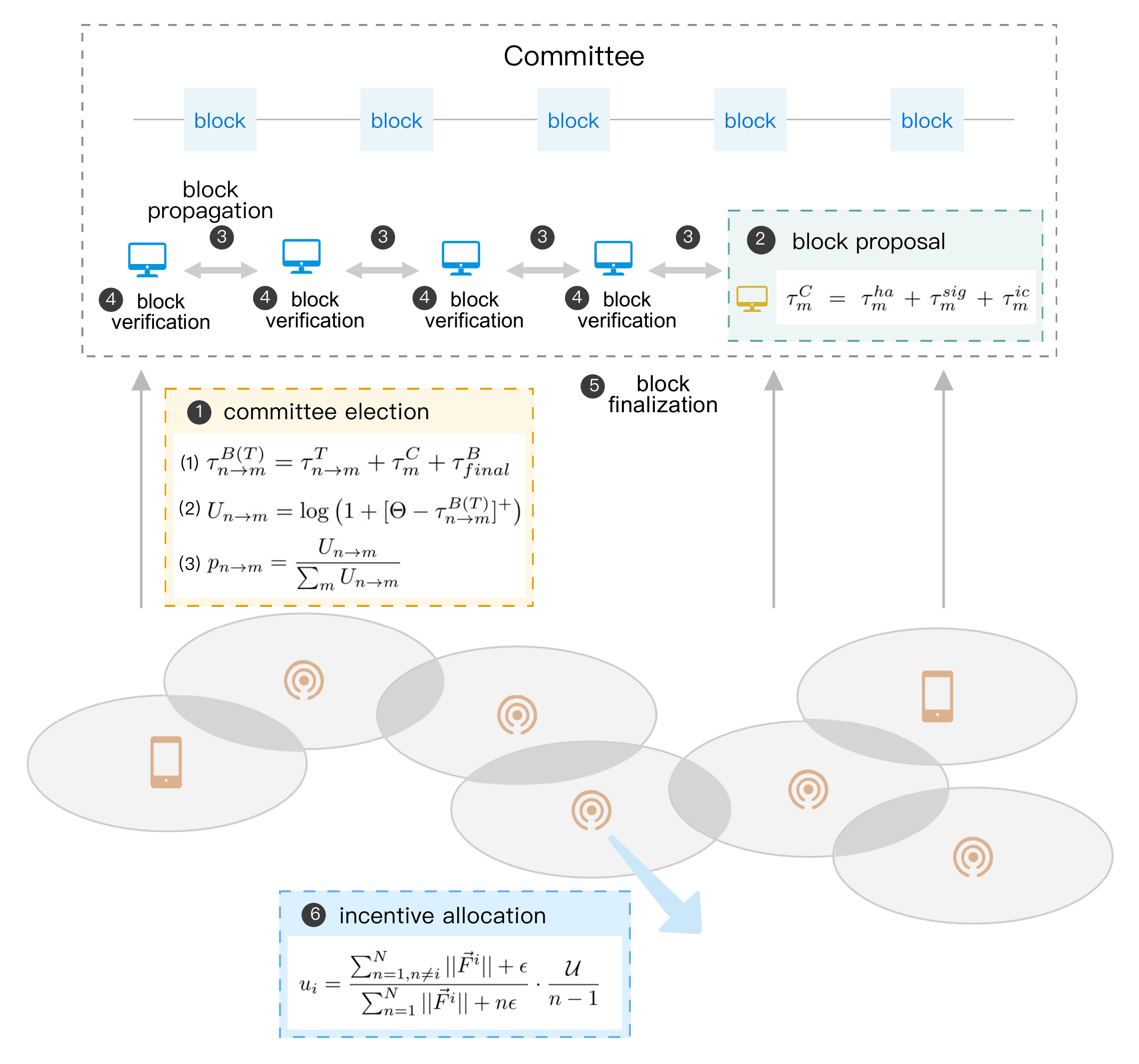}
	\caption{The overview of location based consensus.}
	\label{fig:label11}
\end{figure}

In this subsection, we present the details of the location-based consensus algorithm. Inspired from DPoS and aiming to maximize the monitoring area, the proposed consensus algorithm consists of six parts: 1) committee election, 2) block proposal, 3) block propagation, 4) block validation, 5) block finalization and 6) incentive mechanism. Different from DPoS or another consensus, the participants of Bychain are given a different number of votes according to its last location reward. The reward allocation  (or incentive allocation) scheme is one of the vital consensus procedures for permissionless blockchain \cite{xiao2020survey}, especially to encourage the profit-oriented blockchain participants to act in the way Bychain expect them to. The details about the proposed consensus are introduced in the following.

\subsubsection{Committee Election}  Although every participant in Bychain can be elected as a committee member, the computation capacity limited by hardware should meet the minimum requirement of the block and consensus computation. On the other hand, instead of DPoS, where stockholders vote for a committee candidate according to personal willingness, the IoT devices in Bychain prefer the candidate with high utility, i.e., the higher network quality and higher computation capacity, the higher probability to be voted. 

From the viewpoint of IoT node $n$, the timestamps of transaction generation, block proposal and block finalization can be observed in the finalized blocks. Therefore, for given $T$th transaction that collected by $m$th committee member in block $B$, the delay of transaction and block procedure can be written as,
\begin{equation}
\tau_{n \rightarrow m}^{B(T)} =  \tau_{n \rightarrow m}^{T} + \tau_m^{C} + \tau_{final}^{B},
\end{equation}
where $\tau_{n \rightarrow m}^{T}$ denotes the delay of transaction propagation from node $n$ to committee member $m$. The block computation delay $\tau_m^{C}$ consists of three parts: a) the hash time consumption $\tau_m^{ha}$, b) the signature generation delay $\tau_m^{sig}$, and c) the incentive computation delay $\tau_m^{ic}$, $\tau_m^{C}=\tau_m^{ha}+\tau_m^{sig}+\tau_m^{ic}$. $\tau_m^{ic}$ will be described in next subsection IV-C. $\tau_{final}^{B}$ is the block finalization delay, which is determined according to the overall network quality. For the permissionless blockchain that may be world-widely deployed, $\tau_{final}^{B}$ is set to be 3s or 0.5s in Bitshares or EOSIO, respectively.

In general, $\tau_{n \rightarrow m}^{B(T)}$ implies the network quality and computation capacity of committee member $m$ that observed by IoT node $n$. Consider the maximum block processing delay of being a committee member is $\Theta$, we define an utility function to indicate the election,
\begin{equation}
U_{n \rightarrow m} = \log\big(1+[\Theta-\tau_{n \rightarrow m}^{B(T)}]^{+}\big),
\end{equation}
where $[a]^{+} = \max\{a,0\}$. $U_{n \rightarrow m}$ is set to be 0 when the block processing delay observed by $n$ is less than the given threshold $\Theta$. Otherwise, $U_{n \rightarrow m}$ would achieve a logarithmic increment. Therefore, the probability of $n$ voting for $m$ can be written as,
\begin{equation}
p_{n \rightarrow m} = \frac{U_{n \rightarrow m}}{\sum_{m} U_{n \rightarrow m}},
\end{equation}
which means the higher utility of $m$, the higher chance to be voted. Then the committee $\mathcal{M}=\{1,..,M_{c}\}$ is elected as the top $M_{c}$ winners.

\subsubsection{Block Proposal, Block Propagation and Block Validation} When the committee is elected to take part in a consensus algorithm, at each block generation slot, the committee members take turns to act as a block miner proposing a block in a round-robin fashion. 

In a specific block proposal process, the miner packages collected PoL transactions into an unverified block. The incentive allocation is also performed in this procedure. It is elaborated in the next subsection in IV-C. Then the unverified block is propagated to other committee members for block validation. The delay of information propagation highly depends on committee members' geographical distribution and the block size. The other committee members verify the signatures of the block proposer and transaction generator. If the verification is successful, the block is appended into the local block database. Otherwise, the appended procedure would be forbidden.

\subsubsection{Block Finalization and Incentive Mechanisms} In Bychain, the block finalization follows the Longest-chain rule. Due to network uncertainty and possible malicious participant, the blockchain may be forked into the longest chain and several shorter chains. In this case, the block miner should always propose a block that extends the longest chain. Otherwise, a local database update should be performed ahead of the block proposal.

There are two kinds of incentive mechanisms involved in Bychain: 1) block generation can claim a certain amount of token reward, 2) IoT devices pursuing maximizing monitoring area can be rewarded a certain amount of votes. 1) is commonly applied to encourage more participants to join in the permissionless blockchain systems, e.g., Bitshares, Bitcoins, etc. We will describe 2) in detail.

\subsection{Incentive Allocation for Maximizing Monitoring Area}
In this subsection, we detail the incentive allocation mechanism to maximizing the total coverage area. For every 24 hours, or every 28800 blocks at 3s block production interval \footnote{Due to fixed block broadcast interval, the number of blocks is a widely used dimension to measure time difference in blockchain system.}, Bychain issues a fixed-value budget as total incentive of maximizing network communication coverage usage. With omnidirectional mobility, each IoT node is willing to find and deploy itself to where maximizes the next round reward. We define it as an \textit{Incentive Allocation for Maximizing Monitoring Area} problem.

\textbf{Problem}:  \textit{Given a coverage budget worth $\mathcal{U}$, and $N$ incentive-pursuing noncooperative nodes with isotropic communication module of available radius $R$, how should the budget be allocated so that the resulting configuration is tending to maximize the network communication coverage?}.

Before going further, some underlying constraints need to be clarified:
\begin{itemize}
	\item  Global-class computing may not be feasible due to limited computation and power consumption of IoT devices. Furthermore, localization of devices is not globally available because, as mentioned in the last subsection, location ciphertext $L_{p_i,t}$  cannot be publicly understood.
	\item  Bychain opens for anyone to access, read, send, or receive transactions and blocks. Hence, the problem-solution requires high scalability and adaptivity to contain abrupt attendance or absence.
	\item  The connectivity between Bluetooth devices highly depending on the electromagnetic propagation characteristics of the natural environment. In certain scenarios with high path loss and shadow fading, the signal from sources in different directions is necessary to avoid the "dead" coverage zone.
\end{itemize}

Motivated by potential virtual field in robotic navigation and obstacle avoidance \cite{robert}, Bychain applies artificial virtual potential field to address above incentive allocation for maximizing monitoring area problem. Each witness node in the blockchain system is treated as a charged particle, such that artificial electric fields are constructed in a way that each node is repelled by repulsive force $\vec{F}_r^{(i,j)}$ from
other nodes, thereby forcing the blockchain to spread itself throughout the environment. Bychain then introduces an attractive force  $\vec{F}_a^{(i,j)}$ to gather nearby witness nodes, and keep the repetitive coverage area.

\subsubsection{Potential Field of Single Node}
 According to the relationship between potential field and force, given $i$th witness node and $j$th near-by witness node, the repulsive force with respect to the scalar potential field $P_j$ can be written as,
\begin{equation}\label{6}
\vec{F}_r^{(i,j)} = -\nabla P_j =\left\{ \begin{aligned}
&- k_{rep} \cdot \frac{1}{r_j^2} \cdot \frac{\vec{x}_j-\vec{x}_i}{r_j}, & r_j \leq R_r , \\
& 0, &  r_j > R_r.\\
\end{aligned} 
\right.
\end{equation}
where $k_{rep}$ denotes the strength constant of the repulsive field, ${r_j}$ is the Euclidean distance between node $j$ and node $i$, where $r_j = ||\vec{x}_j-\vec{x}_i||_2$ , $\vec{x}_j$ and $\vec{x}_i$ denote the absolute position of node $j$ and node $i$, respectively. In general, $\vec{F}_r^{(i,j)}$ is a conservative force that subjects to the gradient of potential field. The piecewise function indicates that each node out of the disc of radius $R_r$ would not affected by its repulsive potential field.

 Let us focus on the potential field of $i$th witness node. As illustrated in Fig.4(a), regardless of the attractive force $\vec{F}_a^{(i,j)}$, $\vec{F}_r^{(i,j)}$  could result in non-overlapping network coverage with uncovered gaps. Indeed, the repulsive-only field could perform well as an initial deployment solution of densely placed nodes. However, it probably conduces a discrete dot-like coverage in urban scene. Therefore, a larger range of attractive field is necessary to gether nearby nodes out of the repulsive field.

\begin{figure}[ht]		
	\centering
	\includegraphics[width=3.5in]{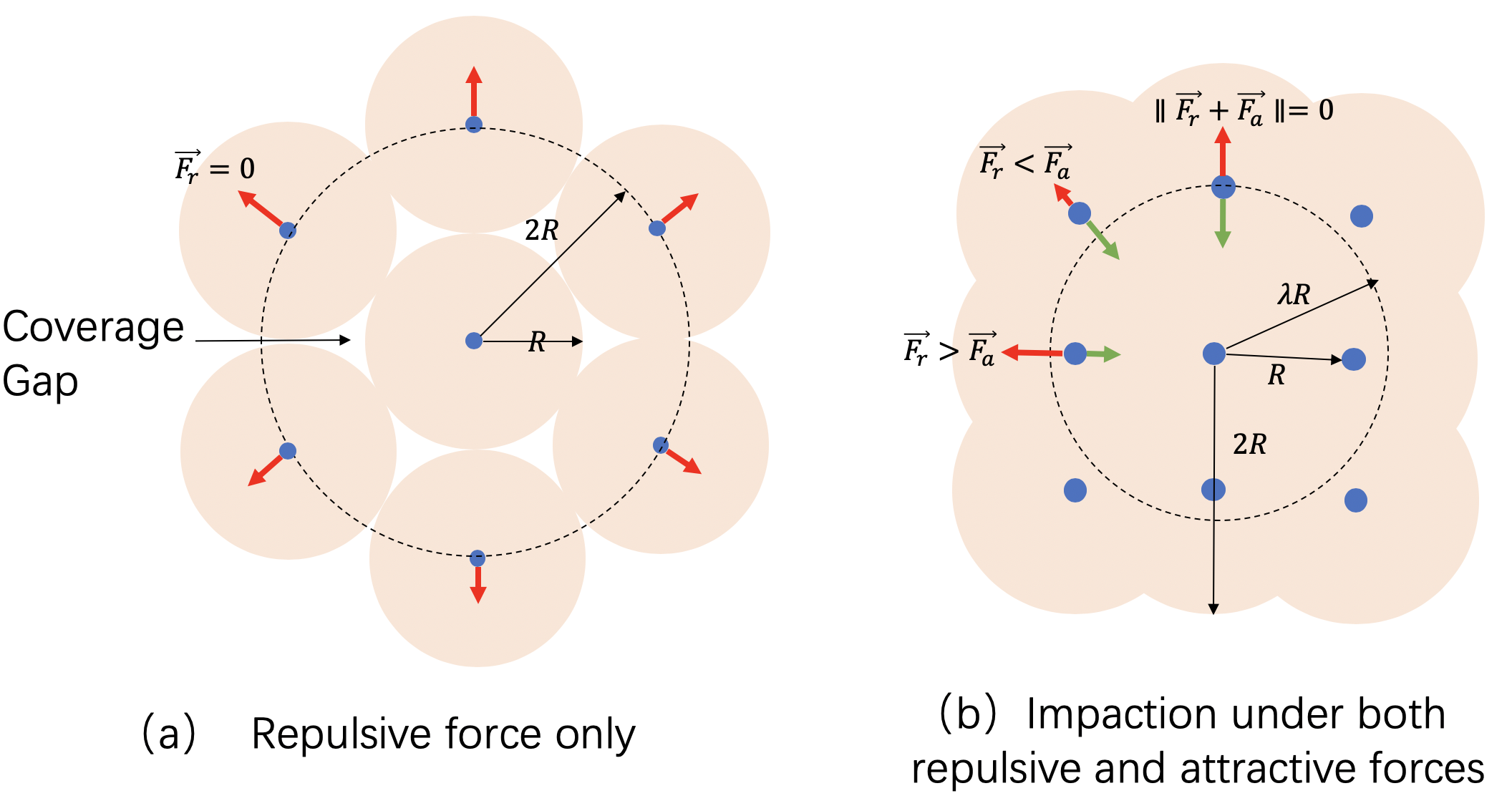}
	\caption{Illustrations of final coverage caused by potential field. In subfigure (a), we set $R_r = 2R$ and remove the effects of attractive force $\vec{F}_a^{(i,j)}$, where dotted circle denotes the range of repulsive field. At this moment the coverage is maximized, however, resulting in several coverage gaps that lifts the interruption probability of Bluetooth connections. After taking attractive force $\vec{F}_a^{(i,j)}$ into account, subfigure (b) showns a node reaches the force equilibrium at $\lambda R$.}
	\label{fig:label4}
\end{figure}

As illustrated in Fig.4(b), an attractive field of $\vec{F}_a^{(i,j)}$ is constructed subjects to $||\vec{F}_a^{(i,j)}+\vec{F}_r^{(i,j)}||=0 $ on the circle of radius $\lambda R$. Furthermore, for nearby node $j$, we wish $\vec{F}_a^{(i,j)}$ growth exponentially with $r_j$, such that node $i$ is able to pull far-distant node back. Hence, $\vec{F}_a^{(i,j)}$ is constructed as following,
\begin{equation}\label{7}
\vec{F}_a^{(i,j)} = \left\{ \begin{aligned}
&- k_{att} \cdot \frac{1}{(r_j-R_r)^2} \cdot \frac{\vec{x}_j-\vec{x}_i}{r_j}, & r_j \geq R_r, \\
& 0, &  r_j < R_r.\\
\end{aligned} 
\right.
\end{equation}
where $k_{att}$ denotes the strength constant of the attractive field. The resultant force between node $i$ and node $j$ is 
\begin{equation}\label{8}
\vec{F}^{(i,j)} = \vec{F}_a^{(i,j)}+\vec{F}_r^{(i,j)}.
\end{equation} Meanwhile, the combined attractive and repulsive potential field construct a zero potential ring at radius $\lambda R$ such that  $||\vec{F}_a^{(i,j)}+\vec{F}_r^{(i,j)}||=0$, shown as the red circle in Fig.5. By substituting \eqref{6}, \eqref{7} and $r_j = \lambda R$ into constraint $||\vec{F}_a^{(i,j)}+\vec{F}_r^{(i,j)}||=0 $, the relationship between strengths of potential fields $k_{rep}$ and $k_{att}$ could be written as following,
\begin{equation}
 k_{att} \cdot \frac{1}{(\lambda-1)^2R_r}- k_{rep} \cdot \frac{1}{\lambda^2 R^2} =0.
\end{equation}
Hence,
\begin{equation}
\lambda^2 k_{att} =  (\lambda-1)^2 k_{rep}  .
\end{equation}

\begin{figure}[ht]		
	\centering
	\includegraphics[width=3.5in]{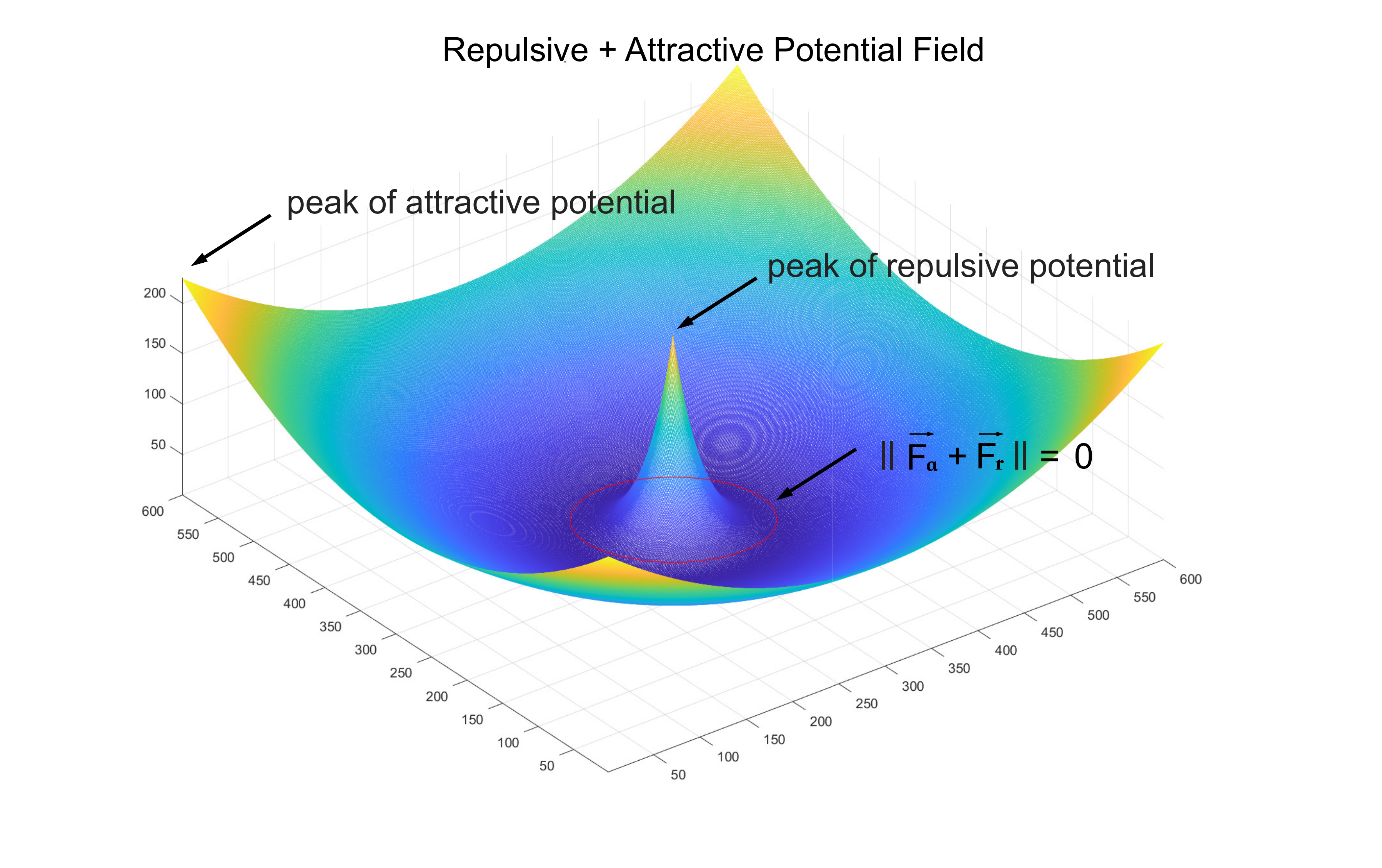}
	\caption{Illustration of the combined potential field, where a witness node locates at (200,200).}
	\label{fig:label5}
\end{figure}

\subsubsection{Potential Field and Intensive of Multiple Nodes}

Let's extend the perspective to the relationship between multiple nearby nodes, denote as a node set  $\mathcal{K}$. The total force of node $i$ can be given by,
\begin{equation}
\vec{F}^{i} = \sum_{j \in \mathcal{K}}  \vec{F}_a^{(i,j)}+ \sum_{j \in \mathcal{K}} \vec{F}_r^{(i,j)}
\end{equation}
The total force of node $i$ results in an acceleration that adjust its velocity  according to the well-known \textit{Equation of Motion}. Therefore, given consensus round $k$, node $i$ with velocity $\vec{v}_k$, the new velocity $\vec{v}_{k+1}$ can be written as:
\begin{equation}
\vec{v}_{k+1} = \vec{v}_{k} + \frac{\vec{F}^{i} - \alpha \vec{v}_{k}}{m},
\end{equation} 
where $\alpha < 1$ denotes viscosity factor and $m$ is the virtual mass of this node.

\textbf{Network Static Equilibrium and Intensive}: It is obvious that the network could asymptotically approach a static equilibrium because the viscous friction term results in monotonically decreasing system energy over time. Hence, one can see a necessary condition for the maximum coverage of the network is the equilibrium on each node. 

Therefore, if blockchain aims to encourage nodes to participate in above coverage maximization, such objective  should design proper budget allocation strategy to intensive motion of each node. Hence, Bychain allocates the coverage intensive to node $i$ as follow, 
\begin{equation}\label{12}
u_i = \frac{\sum_{n=1,n\neq i}^N||\vec{F}^{n}||+\epsilon}{\sum_{n=1}^N ||\vec{F}^{n}||+n\epsilon}\cdot \frac{\mathcal{U}}{n-1}
\end{equation}
where $\epsilon$ is a perturbation factor that sufficiently small preventing denominator to be 0. \eqref{12} performs three key characters: 1) the incentive of node $i$ would decrease as $||\vec{F}^{i}||$ increase. 2)  When the network static equilibrium, each node would be distributed $\frac{1}{n}\mathcal{U}$ budget. 3) The sum of network budget still approximately equal to $\mathcal{U}$.

 %the  resultant absolute value of force of the network can be given by,
%\begin{equation}
%||\vec{F}^{net}|| = \sum_{n=1}^N ||\sum_{j \in \mathcal{K}}  \vec{F}_a^{(i,j)}||+  \sum_{n=1}^N ||\sum_{j \in \mathcal{K}} \vec{F}_r^{(i,j)}||
%\end{equation}

%It's natural to consider that all nodes of force equilibrium satisfies the coverage maximization condition

 %Therefore, a key problem is, \textit{how can we design control low such that $\vec{F}_a$ and $\vec{F}_r$ could maximize coverage while ensuring repetitive coverage?}

%\begin{figure}[ht]		
%	\centering
%	\includegraphics[width=3.5in]{figure11.pdf}
%	\caption{Overview of the Self-Deployment Mechanism.}
%	\label{fig:label6}
%\end{figure}

\section{Security Analysis}\label{sec5}
In this section, we analyze our proposed blockchain's security properties and prove that the protocol can achieve our security goals.

\textit{1) A prover cannot generate a location proof without a witness.}

Since witnesses do not give away their private keys, a prover cannot claim the activities information by signing it using the private key of the witness.  If this situation occurs, the blockchain key cryptography system can detect it.

\textit{2) If the witness does not reveal the private key, a prover cannot generate a legitimate location proof at the claimed location and claimed time without colliding with more than half of the witnesses.}

If a witness does not provide its private key, the digital signature cannot be produced by any node except by the witness itself. In this case, there may be two kinds of attack: (1) the prover may provide a false location claim, or (2) the prover may perform \textit{signature relaying attack}, that is, store the digital signature in the last location and transmit it in the current location.

The attack (1) can be detected easily since the communication between the witness and prover is always near-range communication.
%Editor: Please ensure that the intended meaning has been maintained in the edits in the previous sentence.
The witness will always calculate the distance between the prover and the witness. If the distance calculated from the physical layer is significantly different from the claimed location information, this attack can be easily detected by the witness. (2) There is an auto-increment field $\textrm{nonce}_{w_j}$ corresponding to the $j$th witness $w_j$. Consequently, the $\textrm{nonce}_{w_j}$ is different in each PoL. If the attack (2) occurs, the signature could not be the same due to $\textrm{nonce}_{w_j}$. Therefore, attack (2) can be easily detected by verifying the content.

%\textbf{Proposition 3}:\textit{Without colluding with more than half of the witness, a prover cannot generate a legitimate location proof at the claimed location and claimed time.}

In the proof-of-location stage, the prover may receive more than one proof-of-location response. Since the witness is semi-trusted and majority-honest, the attack can be conveniently detected by analyzing the proof-of-location responses if the prover does not collude with more than half of the witness nodes.

\textit{3) The prover, witness and verifier are not able to modify generated activity proofs.}

In the second stage, the proof-of-location information is uploaded on the blockchain, and the data on the blockchain is transparent and tamper-resistant. It is almost impossible to modify the generated activity proofs.

\textit{4) This protocol is able to avoid the PoL replaying attack on the blockchain.}

Due to the consistency of the blockchain database, the on-chain storage space is precious. A malicious user may resubmit the PoL commitment to occupy the storage space, named the \textit{PoL replaying attack}.
Since each peer of the blockchain network checks, if the current PoL is contained in the blockchain database before submitting, it is not possible to successfully complete the replaying attack.

\textit{5) A verifier can verify the activity information without the private key of the proof}.

The zero-knowledge proof is applied in our protocol. The private key will not be revealed to the service provider. The signature of the activity information is verified by providing the verifier, a signed verifier-specified nonce, and this kind of attack can be avoided.

\textit{6) Only the prover and verifier can access the location information in this protocol, such that the privacy is maximized}.

While these activity proofs are uploaded on the blockchain such that they are transparent to all the other users, the public key and private key are managed by key escrow and changed in each $k$ proof-of-location stage. No one can connect this information with an uniform identity. At the same time, the user does not need to reveal its private key to anyone else when verifying ownership, such that the risk of revealing the private key is avoided. Hence, privacy is maximized.

%\textbf{Proposition 8} \textit{ The real identities of provers is not reveal to anyone else}.

%	\begin{figure}[ht]		
%	\centering
%	\includegraphics[width=3.5in]{figure12.pdf}
%	\caption{Ideal Witeness Coverage Area}
%	\label{fig:label6}
%\end{figure}

\section{Implementation and Evaluation}\label{sec6}

\begin{figure}[ht]		
	\centering
	\includegraphics[width=3.5in]{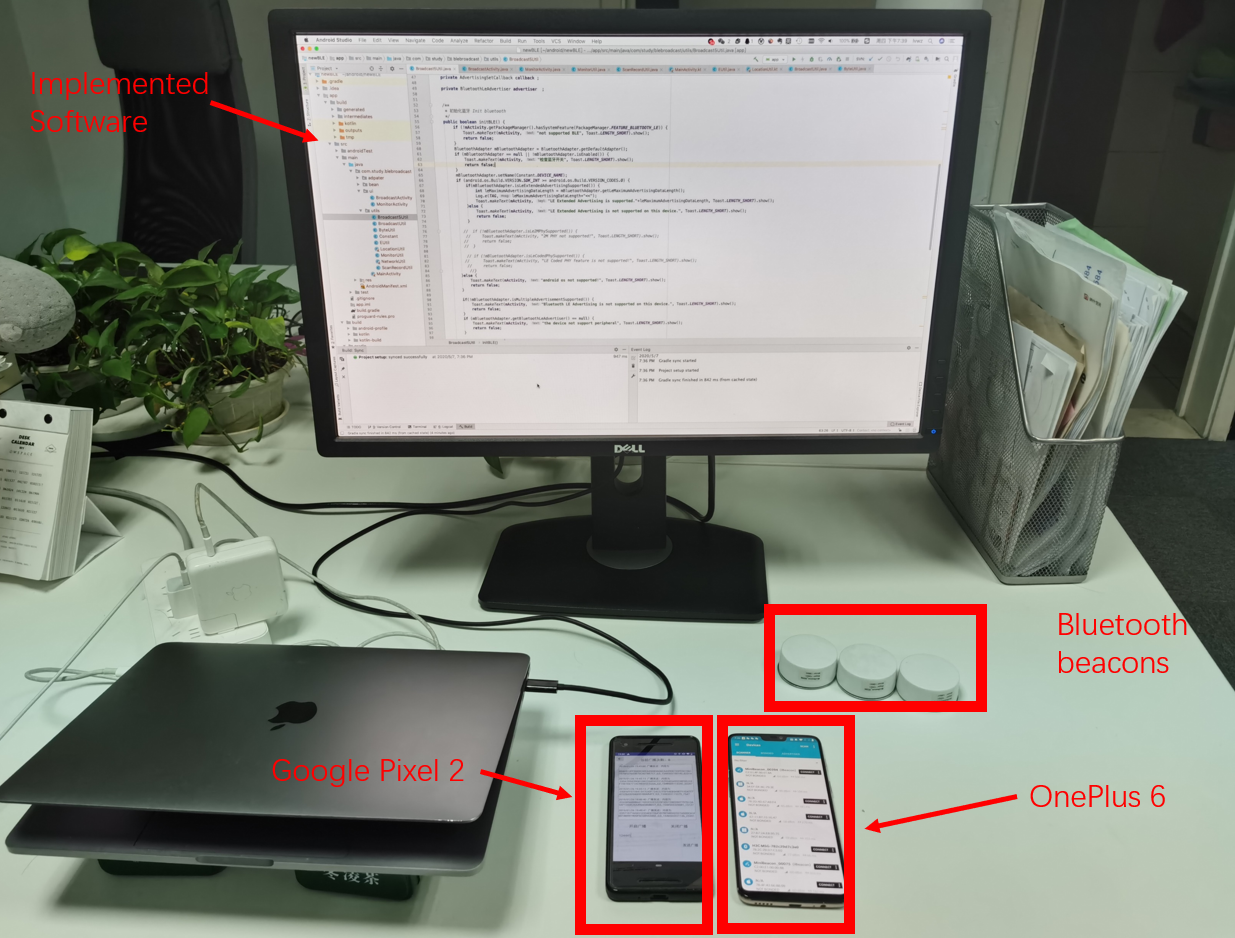}
	\caption{Implemented System which is composed of a Google Pixel2, an One Plus 6, several Bluetooth beacons and the blockchain nodes that running on Alibaba Cloud.}
	\label{fig:label7}
\end{figure}

To verify our system in a real-world setting, we implement production-grade software by using Bluetooth 5.0. The implemented system is shown in Fig.\ref{fig:label7}. In this section, we study the performance and constraints of our proposed protocol, such as the message interaction latency, power consumption, computation and storage limitations. The main software components and data structures are shown as follows.
\begin{itemize}
	\item Programming Language: The client is implemented using Java and Kotlin while the blockchain server and smart contract are implemented using C++. Kotlin is a statically typed language with a type inference suggested as an alternative to Java on the Android platform. We use Kotlin to reduce the programming complexity on the client interface. Due to the low computation cost, C++ is considered to be one of the best choices of our blockchain system.
	\item Bluetooth 5.0: This is latest version of the Bluetooth wireless communication protocol. We adopt Bluetooth 5.0 in our protocol to study the feasibility of our protocol under the latest wireless standards. The PoL message exchange can also benefit from the extended advertising data length such that the signal Bluetooth frame is able to carry our PoL requests and PoL responses. In our software, the Bluetooth 5 advertising extension mode is employed. The primary physical layer parameter is LE 1M and the secondary physical layer is LE 2M.
	\item OpenSSL: This is a general-purpose cryptography library for the transport layer security (TLS) and secure sockets layer (SSL) protocols. The OpenSSL library is a successor of commercial-grade and full-featured cryptography implementations. OpenSSL is required in our protocol to provide signing, signature verification, key pairs generation and key escrow operations.
	\item Bitcoinj-core: This is a Java implementation of the Bitcoin protocol for building blockchain applications. Peer-to-peer communication, network nodes organization and the blockchain object database are implemented and applied in our protocol.
	\item nRF-connect: This is wireless performance analysis software that is produced by Nordic Semiconductor, including  Bluetooth, WiFi and cellular signals. The received signal strength indication (RSSI), packet history,  packet changes, Bluetooth advertising intervals, etc., can be analyzed using nRF-connect in real time and in real-world applications.
\end{itemize}

\textbf{Blockchain network and nodes:} We realize our protocol on Alibaba Cloud. On the blockchain, we generate the genesis block (known as the first block of blockchain) using the unique Merkle hash root at 2017-09-05 12:15:00. After two years of running, it is proved that our DPoS blockchain is stable in practice. We then modified its incentive mechanism from March 2020. Because the on-chain data transaction is not heavy, the current size of our private blockchain database is approximately 2.4 GB and the number of network nodes is approximately 65. Note the number of nodes varies from 60--68 during the experiment. To simulate the real world, our blockchain is implemented on an Intel(R) Xeon(R) CPU E5-2650 v3 platform with 2 GB RAM and 10 Mbps bandwidth for each node is provided by Alibaba Cloud and Amazon Cloud. Nodes in the blockchain network are virtual private servers such that they broadcast and receive block messages via a physical communication link. The mining interval is set as 3\,s.

\begin{table*}
	\caption{CPU and power consumption of each entity}\label{tab1}
	\begin{tabular}{@{}llllllll@{}}
		\toprule
		Property                 & Distance $d_{p-w}$(m)  & \multicolumn{3}{l}{CPU utilization(\%)} & \multicolumn{3}{l}{Power Consumption $P_E$ (mW)} \\ \midrule
		&                                     & minimum     & maximum     & mean    & minimum      & maximum     & mean     \\ \midrule
		\multirow{2}{*}{Prover}  & 3\,m                                  & 2           & 7           & 4       & 356          & 552         & 456      \\
		& 15\,m                                 & 3           & 7           & 4.5     & 468          & 636         & 534      \\ \midrule
		\multirow{2}{*}{Witness} & 3\,m                                  & 0.2         & 0.9         & 0.5     & 296          & 350         & 323      \\
		& 15\,m                                 & 0.4         & 1.1         & 0.6     & 394          & 538         & 432      \\ \midrule
		Blockchain Miner         &                                     & 15            &43             & 30      &              &             &          \\ 
		\bottomrule
	\end{tabular}
\end{table*}
\textbf{The provers and witnesses:} To mimic real-world user activities, we developed a client to realize the functions required by the prover and witness. The client is implemented by Java and Kotlin on One Plus 6 and Google Pixel 2 (all with Android 9.0). One plus 6 equips Qualcomm SDM845 Snapdragon 845 (10\,nm) with 4x2.8\,GHz CPU cores and 4x1.7\,GHz CPU cores, 6 GB RAM, WiFi 802.11 a/b/g/n/ac, Bluetooth 5.0 and a GPS module. Google pixel 2 equips Qualcomm MSM8998 Snapdragon 835 (10\,nm) with 4x2.35 GHz CPU cores and 4x1.9\,GHz CPU cores, 4 GB RAM, WiFi 802.11 a/b/g/n/ac, Bluetooth 5.0 and a GPS module. Both One Plus 6 and Pixel 2 can communicate with blockchain via the WiFi or 4G signal provided by China Mobile.

\textbf{Data structure:} The interacted message involved in our protocol includes the following: the PoL request $P^k$, the PoL response $R^k$, the PoL commitment $\textrm{Com}^k$ and the verification information $N^k$. In our protocol, the length of $P^k$ advertising data is approximately 320 bytes including a 128-bit service UUID and payload data shown in \eqref{1}. The message length is heavily dependent on the key length of the cryptosystem. There always exists a trade-off between the performance and security level when deciding the public key length. Our public key is a 160-bit hash of the SHA256 hashed ECDSA public key and the private key is abstract syntax notation one (ASN.1) coded strings. The signature of the PoL commitment is a 65-bit ECDSA formed signature.  As a comparison, the length of $R^k$ is approximately 540 bytes, and the PoL commitment under a single witness is approximately 520 bytes. The location information is required from the GPS module with 14 bit data and $10^{-7}$ accuracy, i.e., centimeter-level location information.

\subsection{System Performance}

In this subsection, we study the performance and cost of deploying our protocol including the storage, CPU, power consumption, and the latency in each procedure. We also measure the system performances under various relative velocities and number of witnesses.

\subsubsection{Storage, CPU and Power Consumption}

 The running of the client code costs approximately 120 MB of data memory, while the blockchain system takes approximately  3.5 GB of data memory. With 62 blockchain network nodes mining the block, the blocks of uploaded location information generate each 6 seconds. Considering that the peer-to-peer network propagation delay is unstable, we monitor the \textit{ arrived time interval of each block} indicating the timestamp differences of the current arrived block and the last arrived block. The result shows that the minimum time interval is 6532 milliseconds and the maximum time interval is 13245 milliseconds. The arrived time interval depends on the network condition, such as the retransmission rate and the verification delay of the PoS algorithm. Therefore, we believe that a powerful CPU and the adjoining blockchain network node may have a positive effect on the blockchain propagation time.

We monitor the CPU utilization of each prover, the witness nodes and the blockchain nodes. When the prover node is broadcasting and the witness node is monitoring, the CPU consumption is approximately 4 percent and 0.5 percent, respectively. This may be due to the listening Bluetooth signal employing less computation than the emitting Bluetooth signals. Another possible reason is that the cryptosystem involved in the prover processing flow costs more in terms of computation. The CPU utilization of blockchain mining nodes is always above 30 percent. We also realize that the CPU utilization of the network mining node reaches a peak when it is mining the block, in which heavy computations such as authentication and encryption/decryption are involved.

The power consumption of the prover, witness and blockchain is studied as well. We monitor the device power consumption before and after the program launched as $P_{\textrm{before}}$ and $P_{\textrm{after}}$, respectively. Hence the power
consumption of each entity can be computed as $P_{E} =P_{\textrm{after}} - P_{\textrm{before}}$. The experiment is evaluated 50 times for each entity to study its statistical properties, which is shown in Table~\ref{tab1}. For each experiment and each entity, we measure the CPU utilization and power consumption with distances $d_{p-w} = 3$\,m and 15\,m between the prover and witness. It is shown that farther Bluetooth signal interaction causes higher power consumption while computation varies little.

\subsubsection{Delay Evaluation}

We evaluate the time delay of each procedure in a relative static environment, where the prover and witness are immobile. The system parameters are shown in Table~\ref{tab2}. During the evaluation, the PoS blockchain system is maintained by 65 mining nodes. The Pixel 2 is in witness mode and One Plus 6 phone is set in the prover mode with the advertising interval $t_{\textrm{adv}} = 100$\,ms, which is the minimum advertising interval provided in the Android Bluetooth Low Energy API . We set the maximum advertising duration $t_{\textrm{adv}}=1000$\,ms in the implementation.

\begin{table}
	\caption{Parameters used in the delay evaluation}\label{tab2}
	\begin{tabular}{@{}llll@{}}
		\toprule
		Parameters & Description                         & Value & Unit \\ \midrule
	$N$	& Number of Blockchain Mining Nodes   & 65    &      \\
	$d_{p-w}$	& Distance between Prover and Witness & 1     & m    \\
	$t_{\textrm{adv}}$	& Advertising Interval                & 100   & ms   \\
	$P_{tx}$	& Prover and Witness Transmit Power Level                & $-$11   & dBm  \\
		& Primary Physical Layer Parameters   & LE 1M &      \\
		& Secondary Physical Layer Parameters & LE 2M &      \\
	$t_{\textrm{adv}}^{\textrm{MAX}}$	& Maximum Advertising Duration        & 1000  & ms   \\
	$S_{\textrm{block}}^{\textrm{MAX}}$ &Maximum Block Size& 2& MB \\
	$S_{\textrm{COMM}}^{\textrm{MAX}}$&Maximum Commitment Size& 9& KB \\
	\bottomrule
	\end{tabular}
\end{table}

The reason for this setting is to try to minimize the detection delay that is caused by random access methods in the Bluetooth protocol. Ideally, the prover's Bluetooth module broadcasts an advertising frame in each advertising event, while the duration of advertising event is controlled by the advertising interval. However, the scanning window of the witness's Bluetooth module is independent of the prover's advertising event, such that the Bluetooth message interaction succeeds if and only if the witness's scanning window coincides with the prover's advertising event. Because the prover does not share its information with witnesses, the message interaction of the prover and witness is considered to be a random access event. We set the minimum advertising interval to improve the possibility of successive Bluetooth message interactions. Hence, the communication delay of the PoL message interaction will be minimized.

\begin{figure}[ht]
	\includegraphics[width=3.5in]{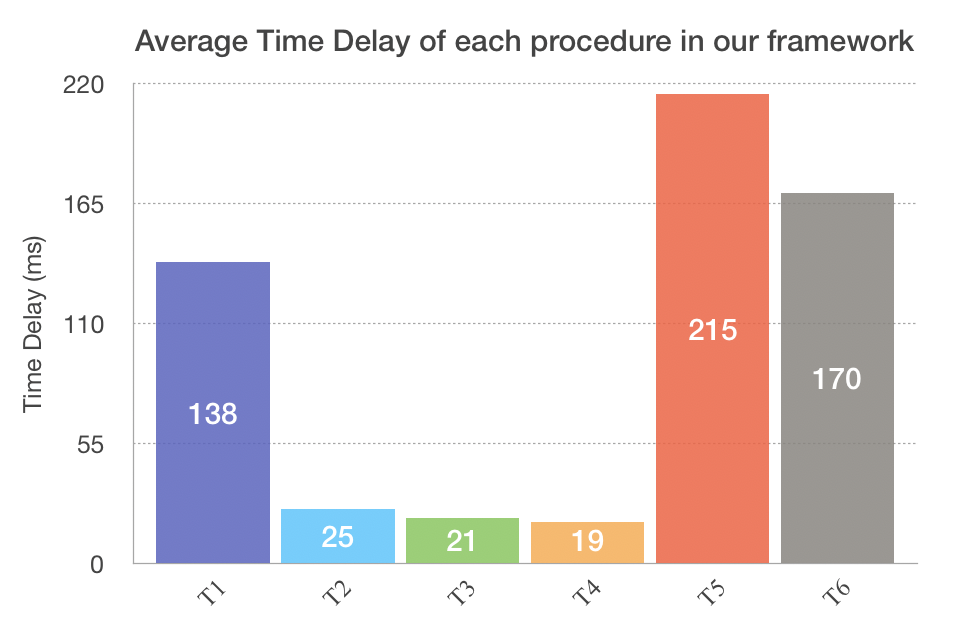}
	\caption{Average time delay of each procedure. T1 is the key escrow generation (key-pair generation for 100 times), T2 denotes the PoL request generation, T3 represents the delay of the PoL request emission task, T4 is the time cost of signature generation, T5 is the time cost of PoL commitment uploading, T6 is the signal propagation delay between the prover and witness. }
	\label{fig:label8}
\end{figure}

We also constrain the blockchain parameters' maximum block size $S_{\textrm{block}}^{\textrm{MAX}}$ and maximum commitment size $S_{\textrm{COMM}}^{\textrm{MAX}}$ as 2 MB and 9 KB to mimic real-world network conditions, respectively. In our experiment, the adoption of Alibaba Cloud and Amazon Cloud may cause unrealistic network conditions because the blockchain miner is considered to be worldwide and under various network conditions. The network condition of our miner nodes are much better than the average blockchain miners such as Bitcoin miners, including the average packet loss probability, the network bandwidth and average latency. A large block size ($>$2 MB) and commitment size ($>$9 KB) increase the computation complexity and bandwidth requirement such that it is much more difficult to maintain the data consistency of the distributed blockchain database. According to the running parameters of the Bitcoin blockchain, we adopt $S_{\textrm{block}}^{\textrm{MAX}}= 2$\,MB and $S_{\textrm{COMM}}^{\textrm{MAX}} = 9$\,KB to mimic a world-wide deployed blockchain system.

To study the sensitivity of our protocol to the time delay of each procedure, we evaluate the time durations of key escrow generation, PoL request generation, PoL request emission, signature generation, PoL commitment uploading and the over-the-air signal propagation delay. After a 50-turn test, the average time delay of each procedure is illustrated in Fig.~\ref{fig:label8}. Among these procedures, the key escrow generation, PoL commitment uploading and the over-the-air signal propagation delay are significantly higher than the others. The PoL commitment uploading is limited by the propagation delay and churn of the peer-to-peer blockchain network; thus it is observed to have high volatility that increases to 421\,ms and decreases to 121\,ms. The signal propagation delay is determined by the channel states of Bluetooth with respect to the relative velocity and activity between the prover and witness. We will further study the influence of key escrow generation and activity in next subsection.

\begin{figure}[ht]
	
	\centering
	\includegraphics[width=3.5in]{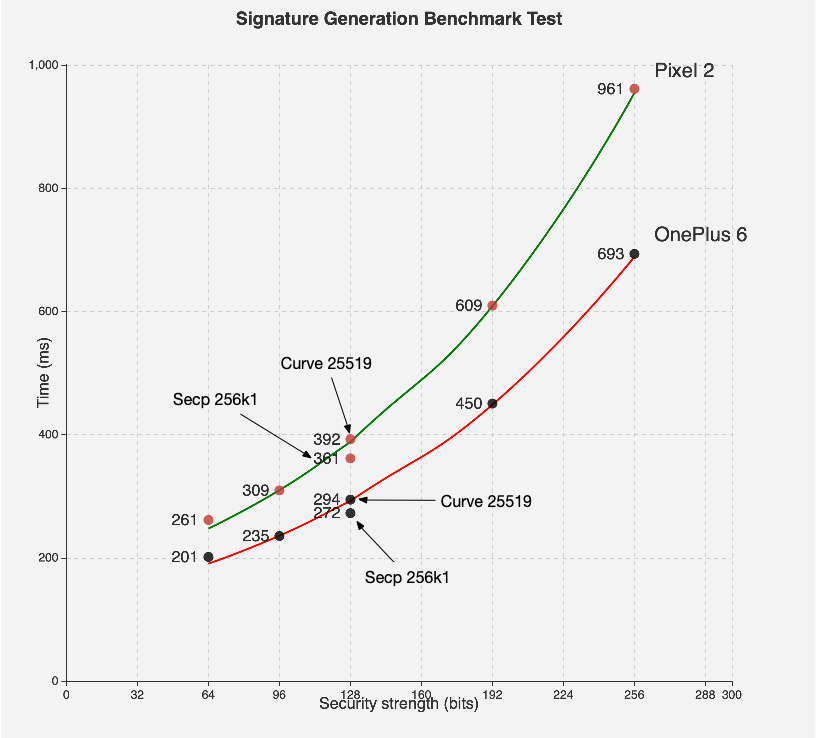}
	\caption{Time comparison under different security strengths. }
	\label{fig:label9}
\end{figure}

The generation of key pairs and signature is studied to evaluate the authentication performance and security. The delay of the key and signature generation and verification is highly dependent on the device performance. We run key pair generation and signature verification 10,000 times regarding secp256k1 (with 128 bits of security strength), secp128r1 (with 64 bits security strength), secp192k1 (with 96 bits security strength), secp384r1 ((with 192 bits security strength)), secp521r1 (with 256 bits security strength), and curve25519 (with 128 bits security strength) on One Plus 6 and Pixel 2. The performances of key generation under different security strengths are shown in Fig.~\ref{fig:label9}. The performance evaluation shows that the time cost increase exponentially with the security strength. However, the curve 25519 algorithm performs with more complexity than secp256k1 even in same security strength. Hence, secp256k1 is implemented in our blockchain system.

\subsubsection{Activities Evaluation}
\begin{figure*}[htbp]
	\centering
	\subfigure[$v=0, a=0,d_{p-w}=1\,\textrm{m}, N=1$ ]{
		\includegraphics[height=1in, width=2in]{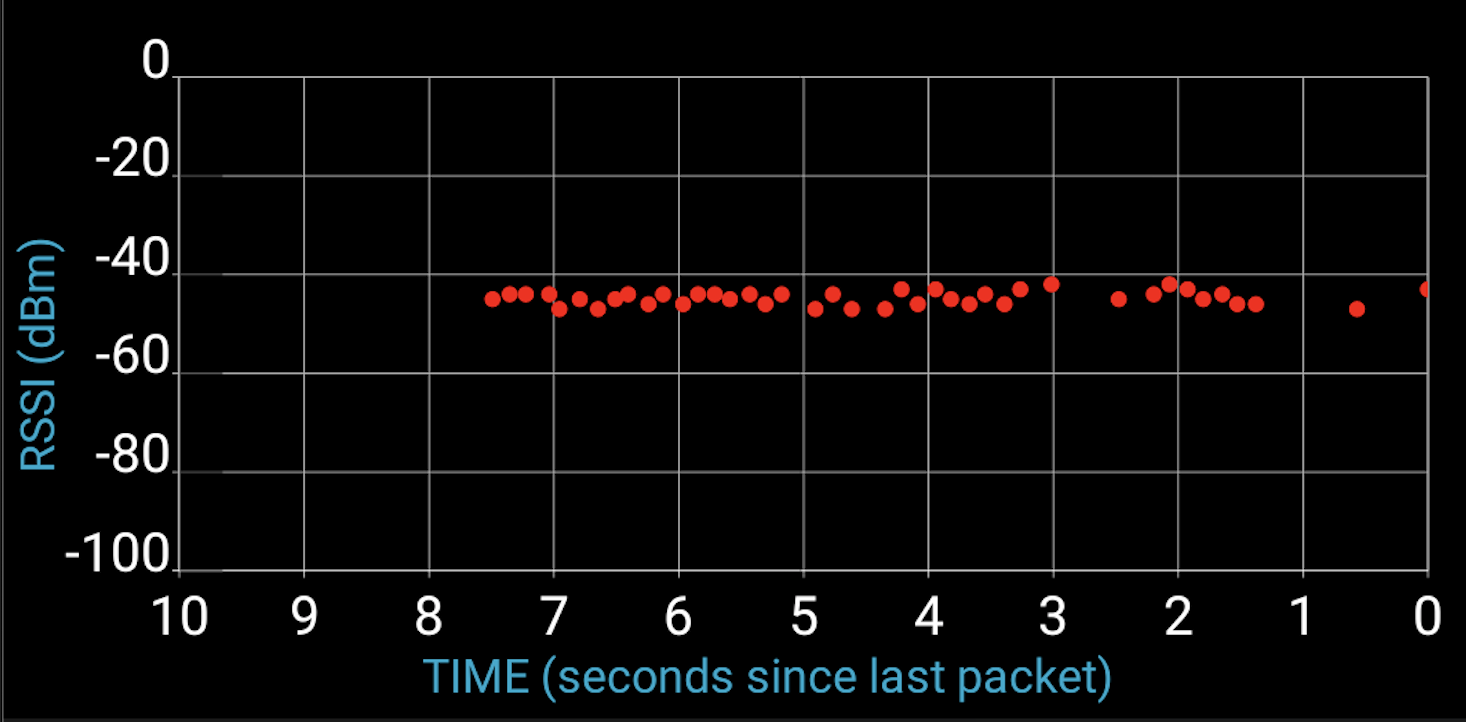}
	}
	\subfigure[$d_{p-w}=3$\,m]{
		\includegraphics[height=1in, width=2in]{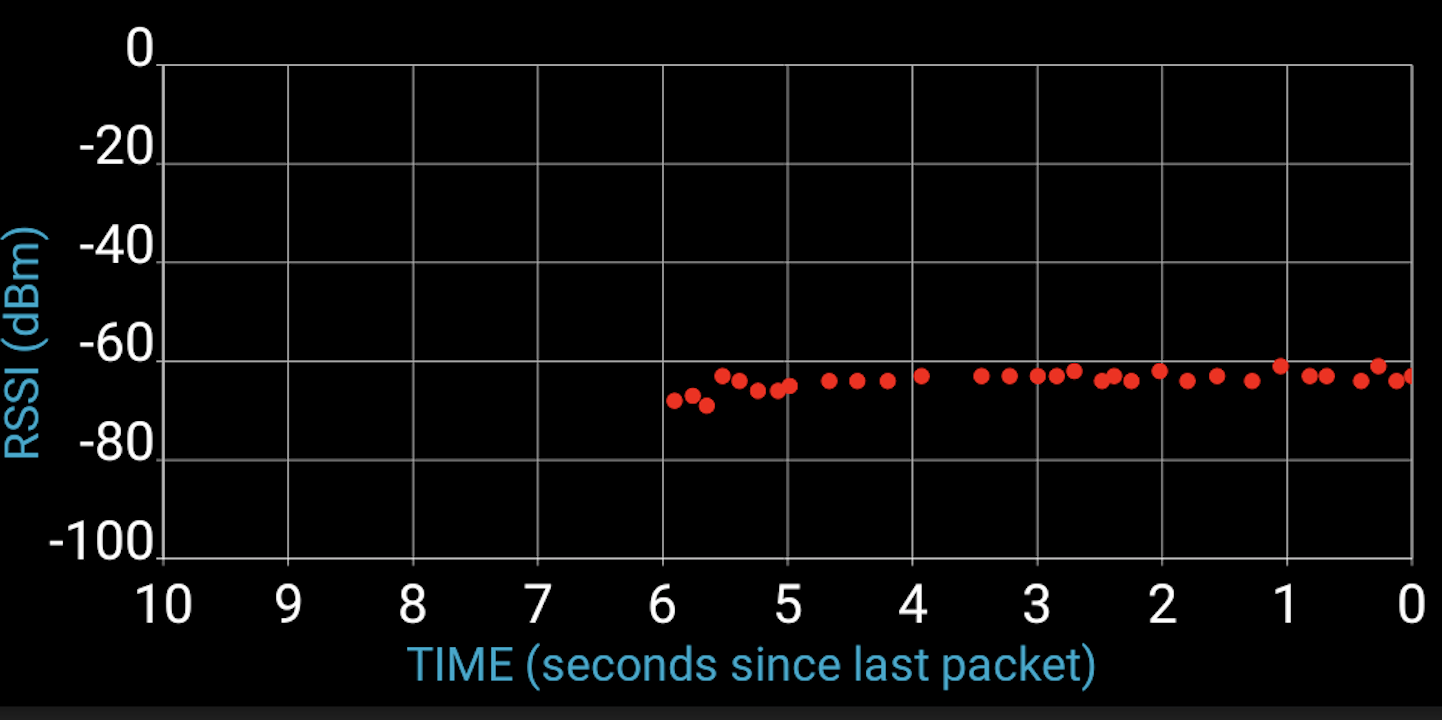}
	}
	\subfigure[$v\simeq 1\,\textrm{m/s}, d_{p-w}=1\,\textrm{m}$]{
		\includegraphics[height=1in, width=2in]{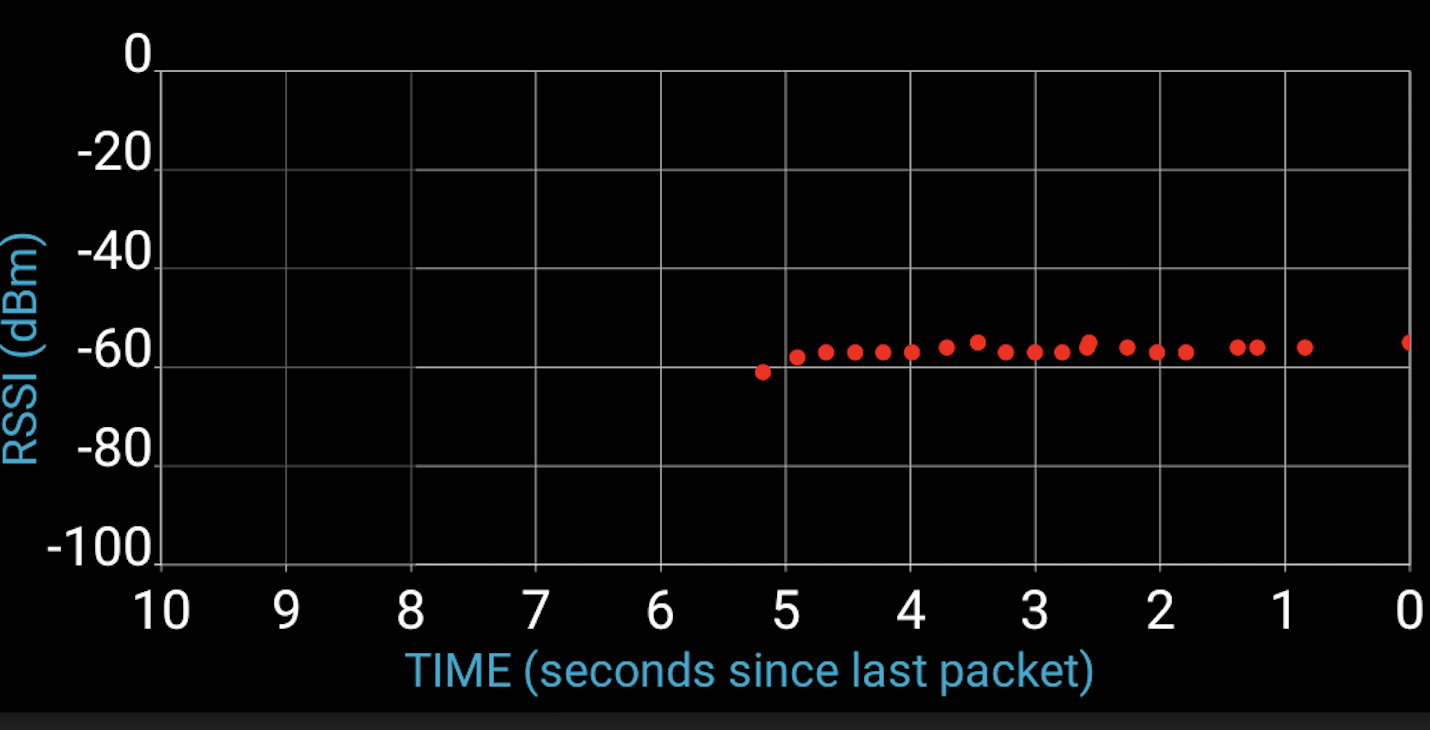}
	}
	\quad    %用 \quad 来换行
	\subfigure[$d_{p-w}=9$\,m]{
		\includegraphics[height=1in, width=2in]{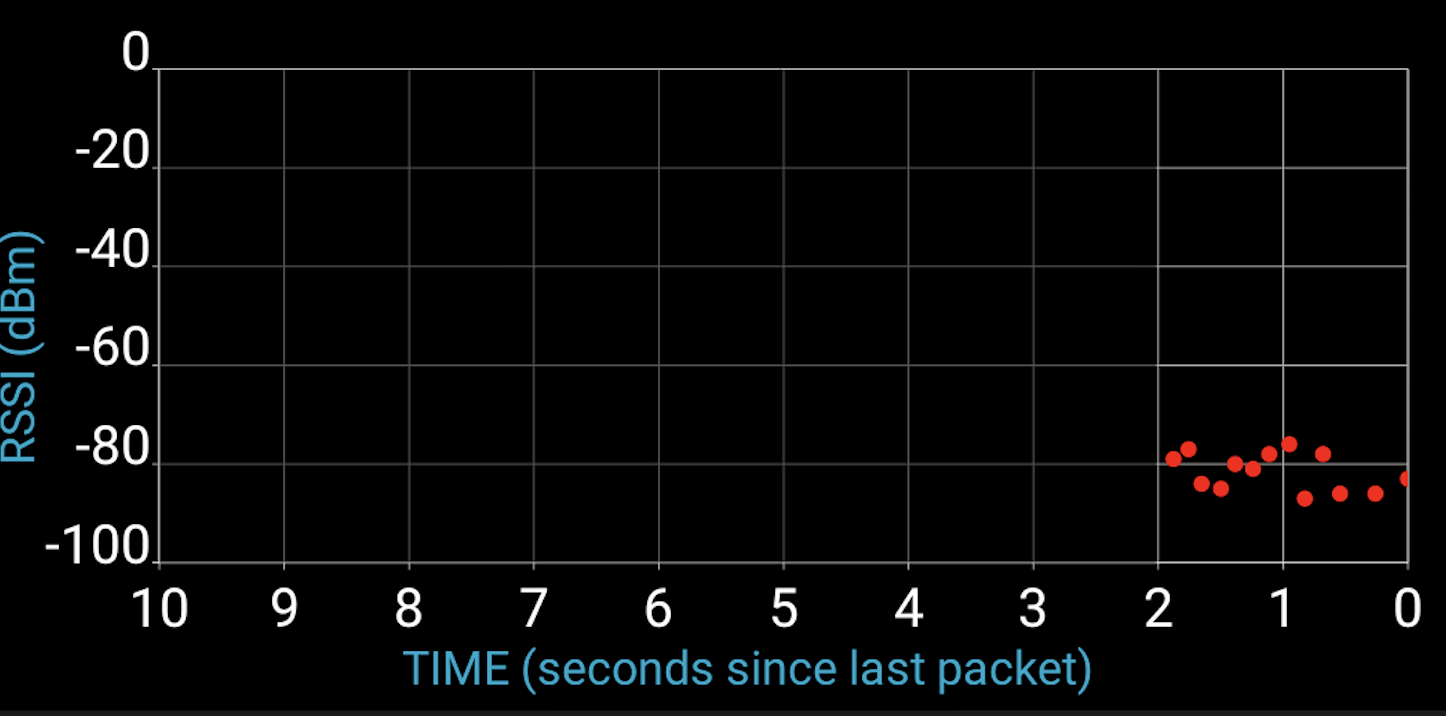}
	}
	\subfigure[$d_{p-w}=9$\,m]{
		\includegraphics[height=1in, width=2in]{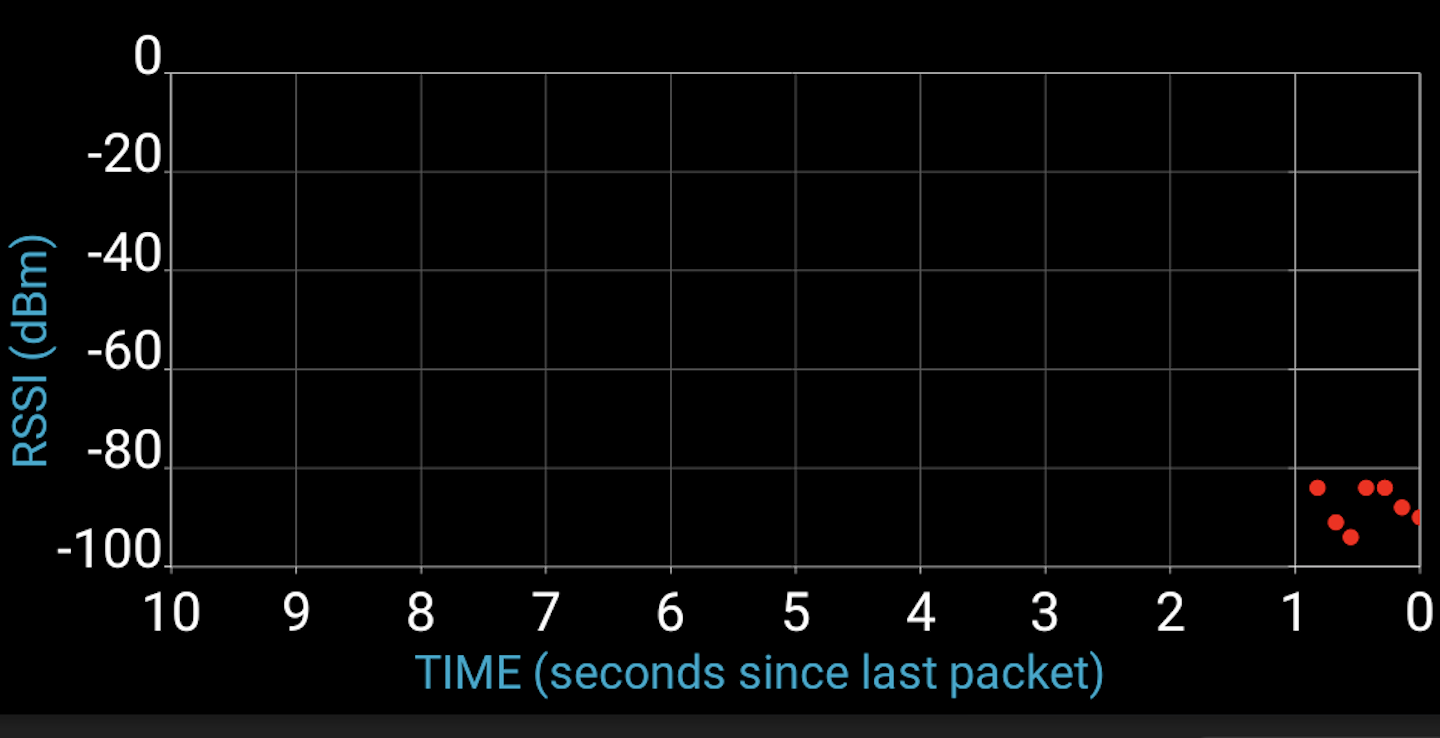}
	}
	\subfigure[$d_{p-w}=3\,\textrm{m},v\simeq 3\,\textrm{m/s}, a\simeq 1\,\textrm{m/s}^2 $]{
	\includegraphics[height=1in, width=2in]{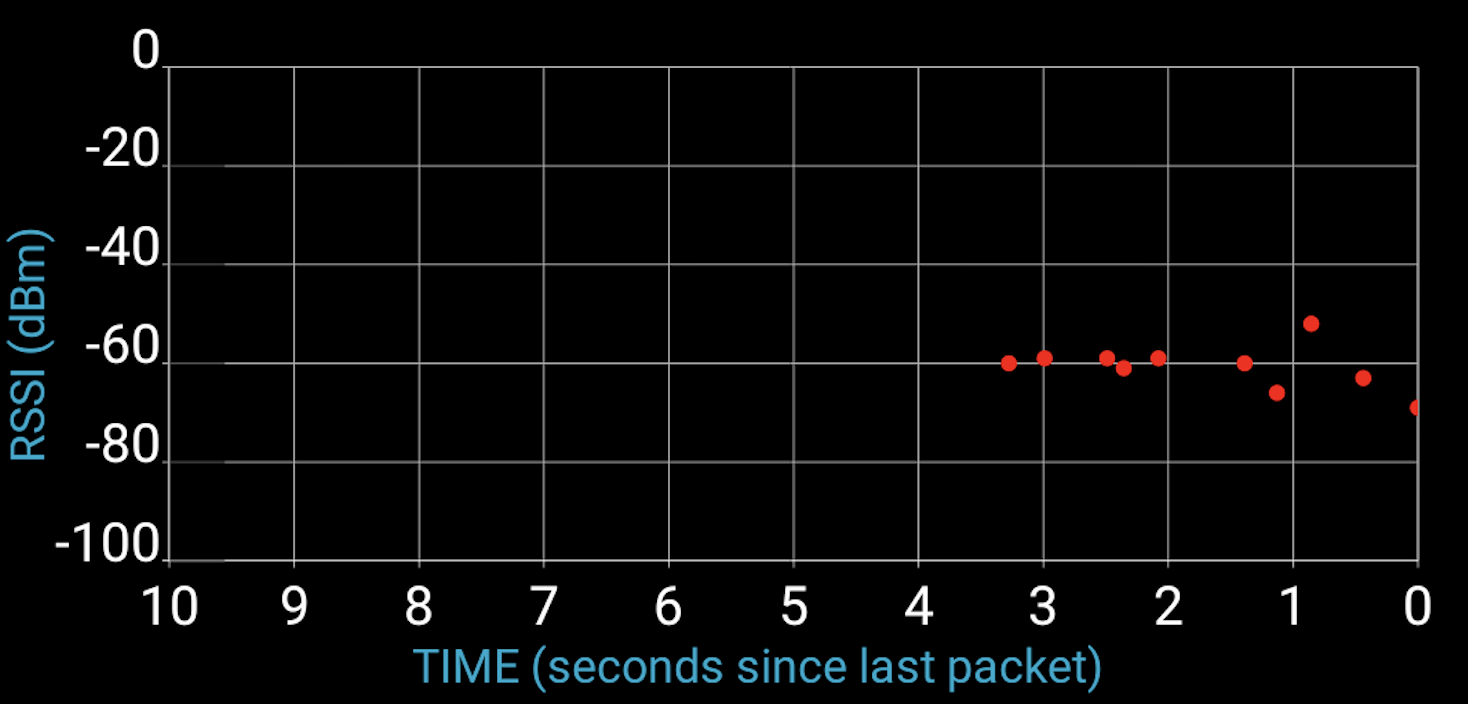}
}
	\quad    %用 \quad 来换行
	\subfigure[$d_{p-w}=3\,\textrm{m},v\simeq 3\,\textrm{m/s}, a\simeq 2\,\textrm{m/s}^2 $]{
		\includegraphics[height=1in, width=2in]{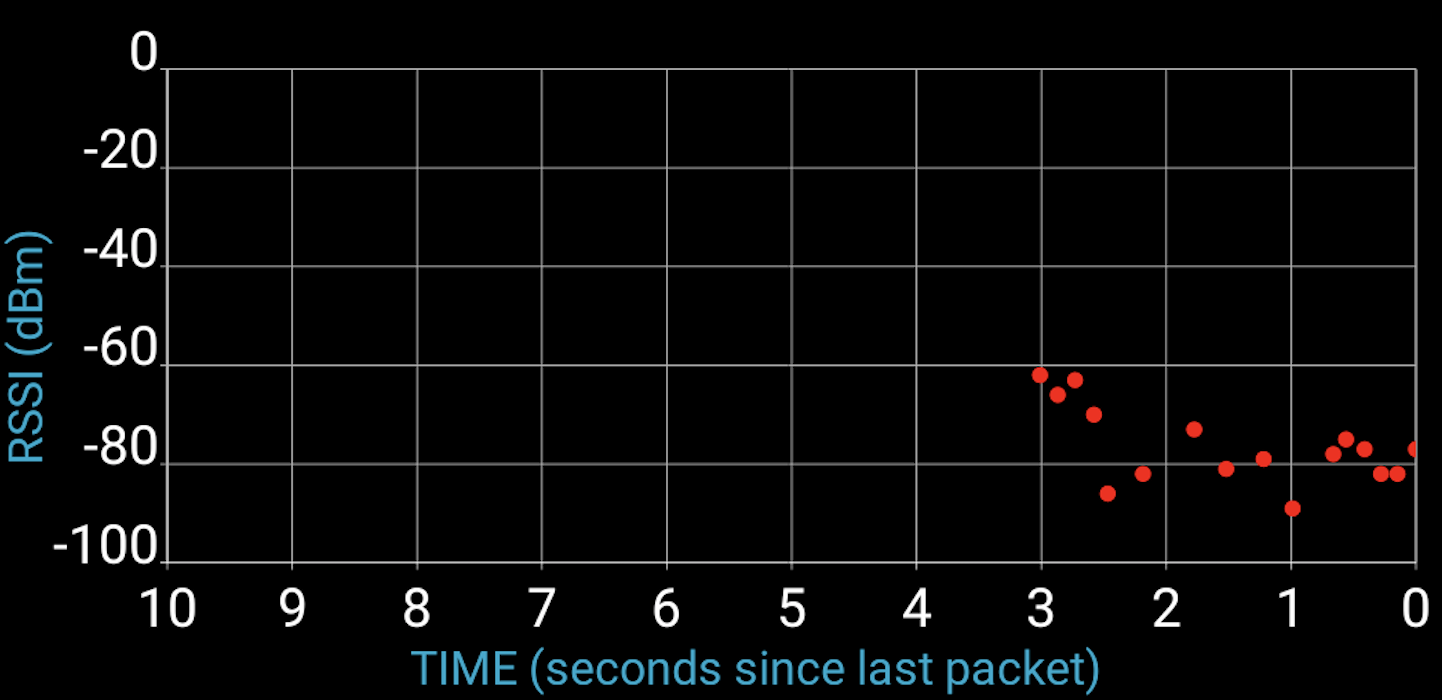}
	}
	\subfigure[Walking through a barrier wall]{
		\includegraphics[height=1in, width=2in]{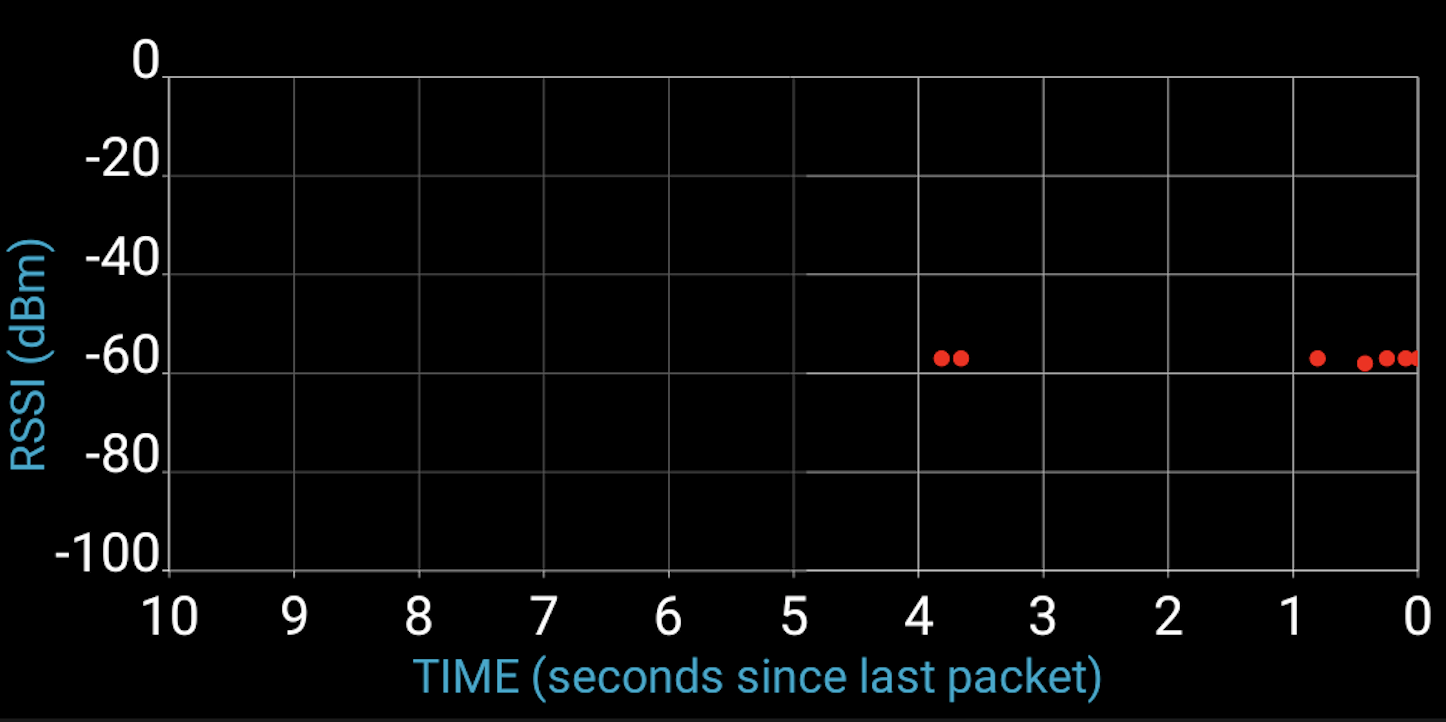}
	}
	\subfigure[Varied direction running]{
	\includegraphics[height=1in, width=2in]{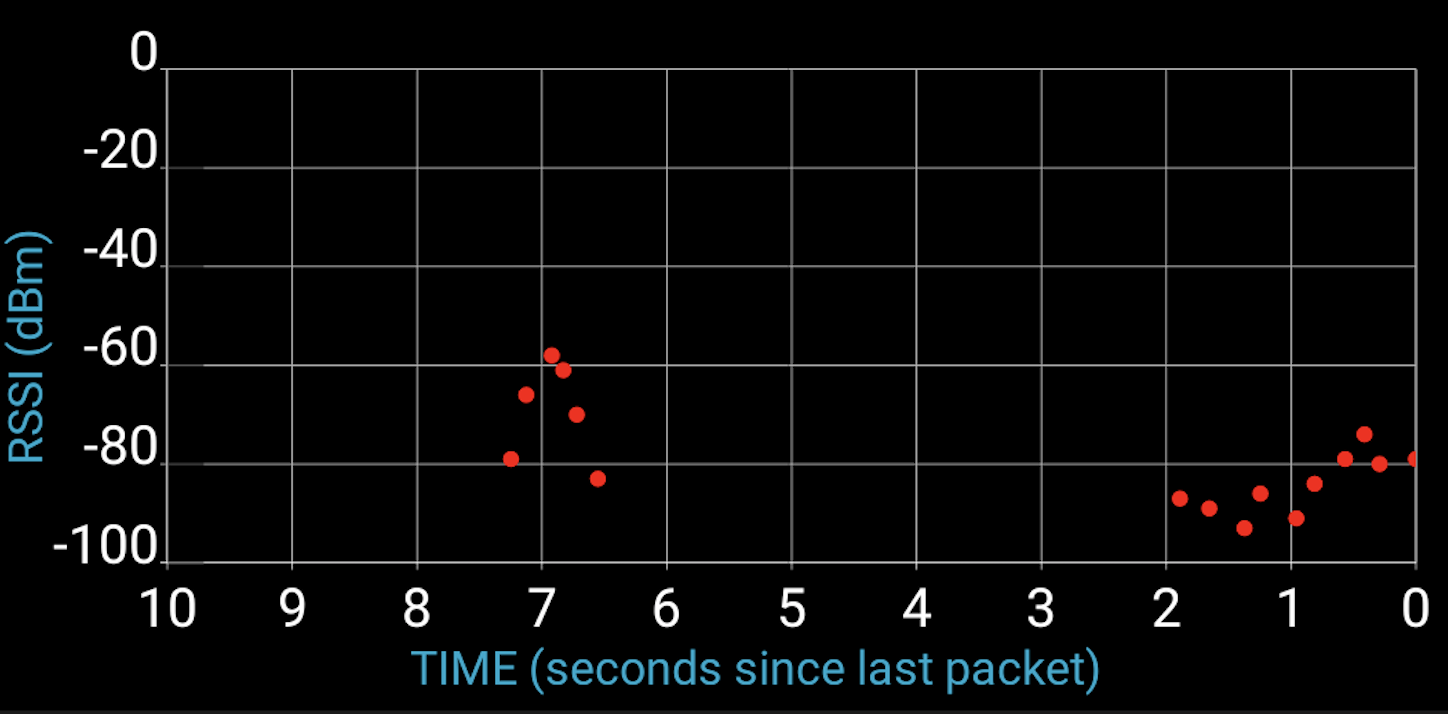}
}
	\quad    %用 \quad 来换行
\subfigure[$d_{p-w}\simeq 7$\,m, $N=5$]{
	\includegraphics[height=1in, width=2in]{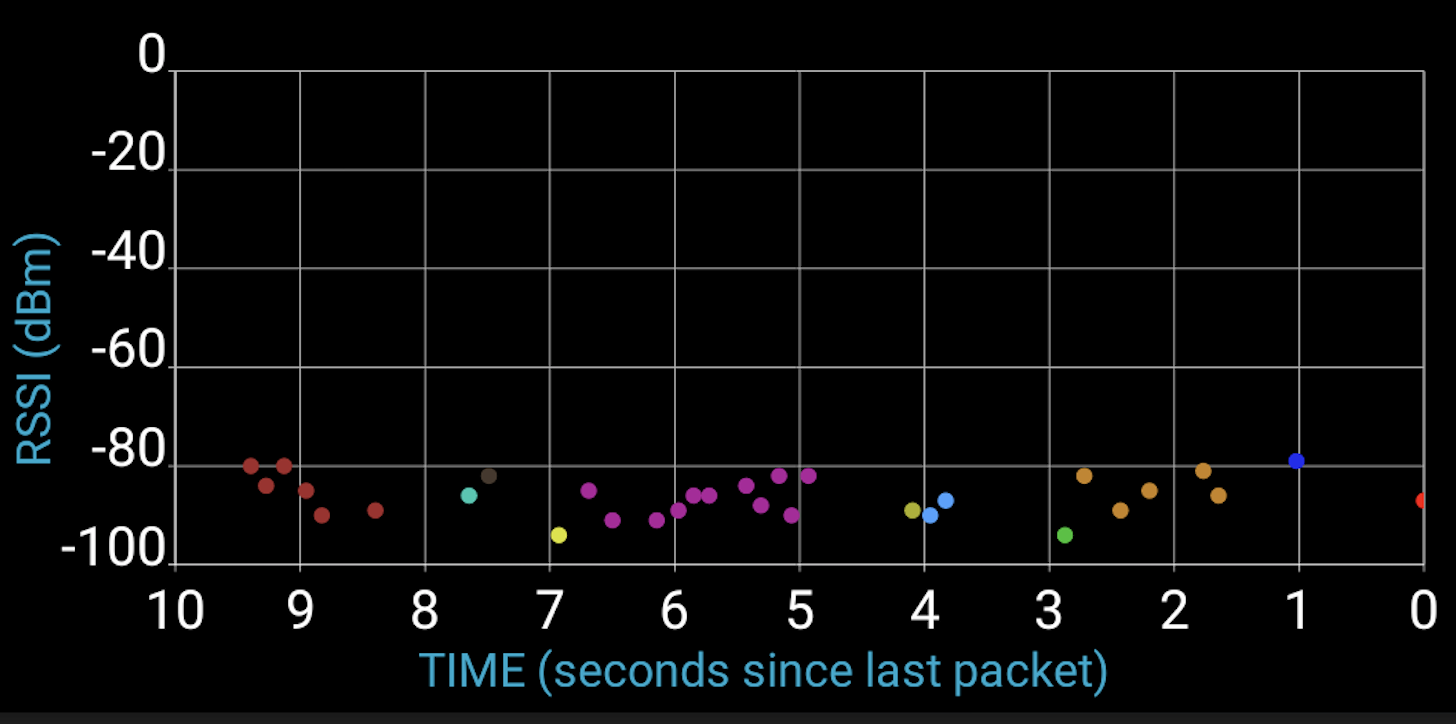}
}
\subfigure[$v\simeq 1$\,m/s, $d_{p-w}=7$\,m, $N=4$]{
	\includegraphics[height=1in, width=2in]{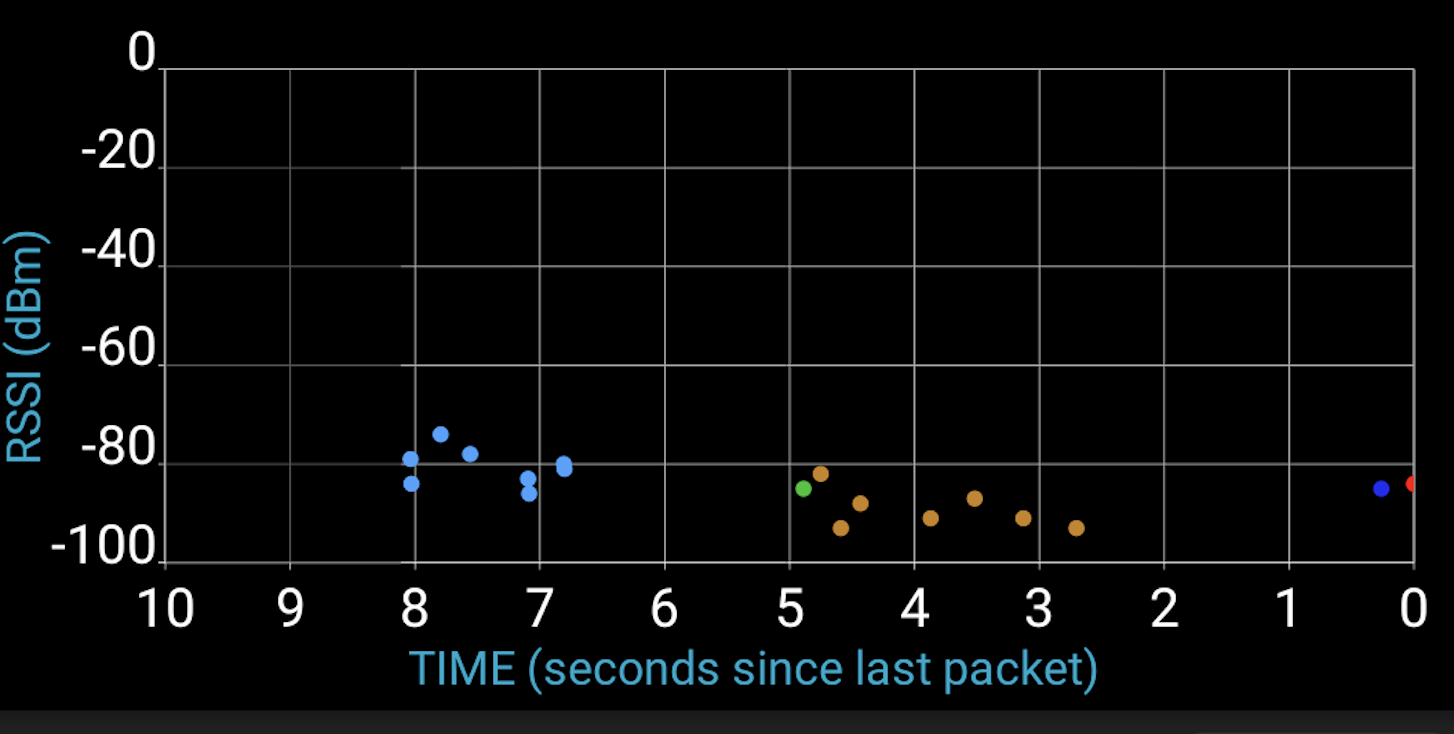}
}
\subfigure[$v \simeq 3\,\textrm{m/s}, d_{p-w}=7\,\textrm{m}, N=7$]{
	\includegraphics[height=1in, width=2in]{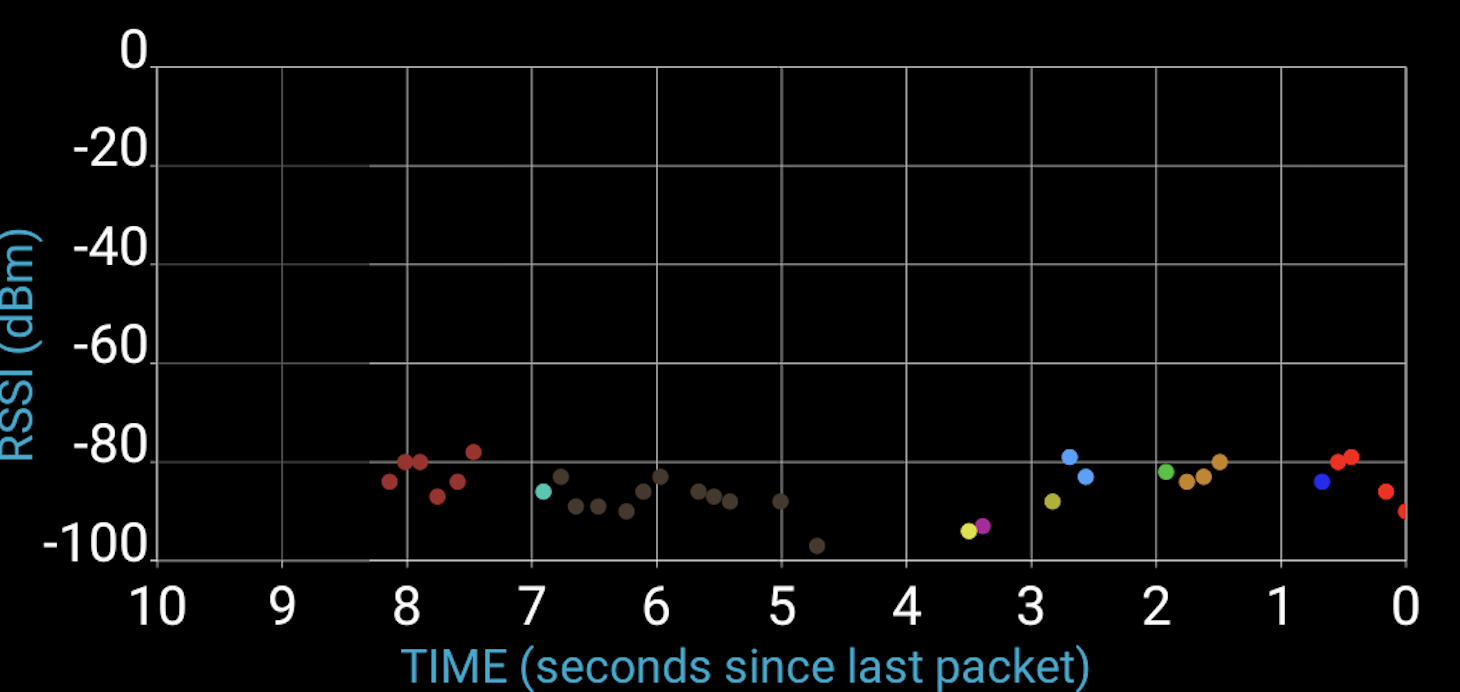}
}
	\caption{Prover Activities Experiment under different velocity $v$, accelerated velocity $a$, initial distance $d_{p-w}$, environment and witness number $N$.}
	\label{fig:label10}
\end{figure*}

%Formatter's Note: Per the journal guidelines, all figures must be cited in the text. Currently, Figure 6 is not cited in the text. Please add a citation for this Figure or remove it from your manuscript and renumber the remaining Figures.

We further evaluate the sensitivity of over-the-air signal propagation to the received signal strength indication (RSSI) and various activity conditions. We adopt Pixel 2 as the prover, all witnesses are static, the experimental tests are conducted in an urban area, and the initial activity condition is shown in Fig.~\ref{fig:label10}(a). It is shown that the RSSI decreases with the more significant initial distance  $d_{p-w}$ and the tremendous accelerated velocity of $a$; this confirms our intuition and suggestions. When a barrier wall appears in the experiment, the Bluetooth signal faces significant fading such that there appears a 2-second gap in subfigure (h). The PoL message interactions are impossible in this signal gap. A prover
walking through multiple witness nodes is evaluated in (j),(k), and (l): the witnesses are placed around the prover in (j), and one-by-one in (k) and (l). The signals that originate from different witnesses are painted with different colors. The result confirms that our protocol works well to collect multiple witness messages in an urban area. Although the PoL message interaction could not coincide, the time interval between each PoL message interaction event is still acceptable at 0.4--2\,s, such that a running user could not run out of the coverage area of Bluetooth 5.0 equipment.

\section{Conclusion}\label{sec7}

In this paper, we proposed and implemented a decentralized blockchain system for activity-proofing and contact tracing.  In the proposed protocol, activity and location-proof problems are addressed by a combination of cryptographic techniques and a decentralized blockchain system without reliance on trusted third parties. The identity privacy problem is protected by proposing a combination technique of zero-knowledge proof and key escrow. The connection of unique cryptographic identity and on-chain proof-of-location commitment is decoupled such that it is almost impossible to track and identify the owner of transparent on-chain data. Hence, security and identity privacy is protected. A location-based consensus algorithm is proposed aiming to maximize the monitoring area. The incentive allocation technology is performed by virtual potential fields. Several implement results are shown to provide a practical perspective.

%Formatter's Note: Your references and in-text citations have been formatted to conform to the journal's guidelines. We encourage you to keep these changes.

\begin{IEEEbiography}[{\includegraphics[width=1in,height=1.25in,clip,keepaspectratio]{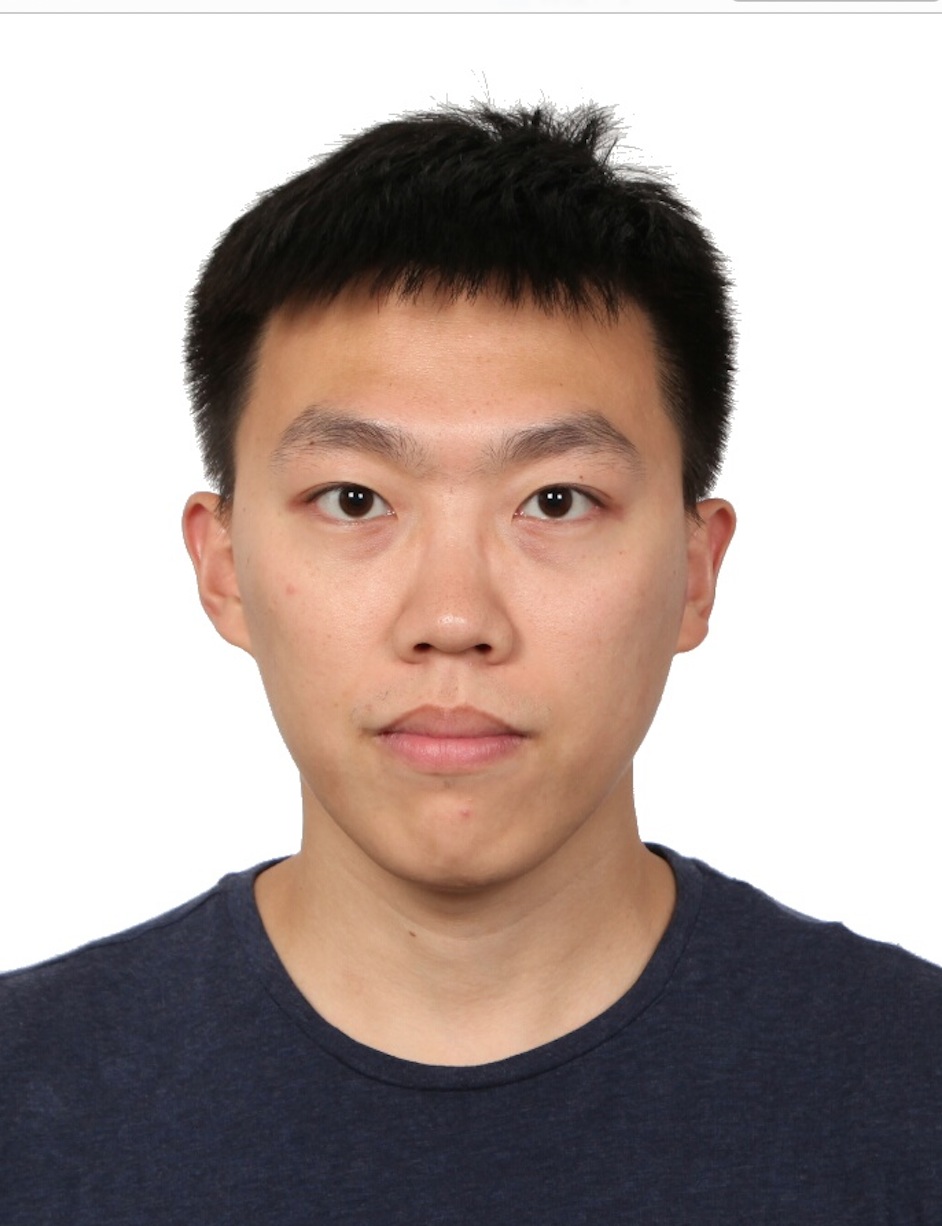}}]{Wenzhe Lv}
	received his B.S degree in Harbin Institute of Technology in 2012. He is currently working toward the Ph.D degree at  Beijing University of Posts and Telecommunications. His research interest focuses on  blockchain technology and distributed system. 
\end{IEEEbiography}

\begin{IEEEbiography}[{\includegraphics[width=1in,height=1.25in,clip,keepaspectratio]{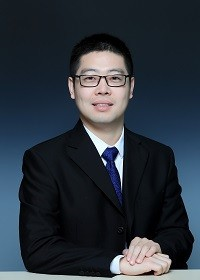}}]{Chunxiao Jiang}
	(S’09-M’13-SM’15) is an Associate Professor in School of Information Science and Technology, Tsinghua University. He received the B.S. degree in information engineering from Beihang University, Beijing in 2008 and the Ph.D. degree in electronic engineering from Tsinghua University, Beijing in 2013, both with the highest honors. His research interests include application of game theory, optimization, and statistical theories to communication, networking, and resource allocation problems, in particular space networks and heterogeneous networks.
	Dr. Jiang has served as an Editor of IEEE Internet of Things Journal, IEEE Network, IEEE Communications Letters, and a Guest Editor of IEEE Communications Magazine and IEEE Transactions on Network Science and Engineering. He has also served as a member of the technical program committee as well as the Symposium Chair for a number of international conferences, including IEEE ICC 2018 Symposium Co-Chair, IWCMC 2018/2019 Symposium Chair, WiMob 2018 Publicity Chair, ICCC 2018 Workshop Co-Chair, and ICC 2017 Workshop Co-Chair. Dr. Jiang is the recipient of the Best Paper Award from IEEE GLOBECOM in 2013, the Best Student Paper Award from IEEE GlobalSIP in 2015, IEEE Communications Society Young Author Best Paper Award in 2017, the Best Paper Award IWCMC in 2017, IEEE ComSoc TC Best Journal Paper Award of the IEEE ComSoc TC on Green Communications \& Computing 2018, IEEE ComSoc TC Best Journal Paper Award of the IEEE ComSoc TC on Communications Systems Integration and Modeling 2018, the Best Paper Award ICC 2019.
\end{IEEEbiography}

\begin{IEEEbiography}[{\includegraphics[width=1in,height=1.25in,clip,keepaspectratio]{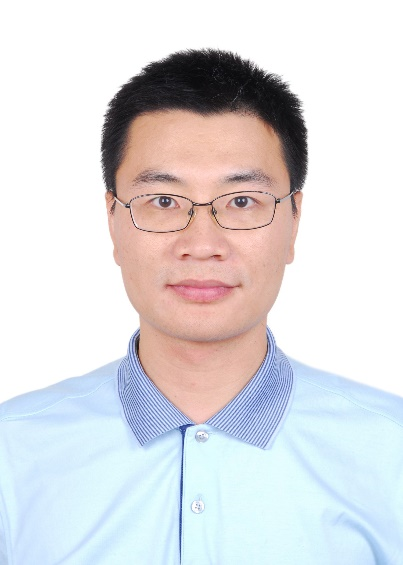}}]{Sheng Wu}
   (S’13-M’14) received the B.S. and M.S. degree from the Beijing University of Posts and Telecommunications, Beijing, China, in 2004 and 2007, respectively, and the Ph.D. degree in electronic engineering from Tsinghua University, Beijing, in 2014. He was a post-doctoral researcher in the Tsinghua Space Center at Tsinghua University, Beijing, China. He is currently an Associate Professor with the Beijing University of Posts and Telecommunications. He has published over 50 journal and conference papers. He also holds over 20 granted patents. His research interests are mainly in iterative detection and decoding, channel estimation, massive MIMO, and satellite communications. He has received the first prize from the Science and Technology Award of Chinese Institute of Electronics in 2017, the silver medal from the 46th Geneva International Exhibition of Inventions in 2018, and the second prize from the National Technological Invention Award of China in 2018.
\end{IEEEbiography}

\begin{IEEEbiography}[{\includegraphics[width=1in,height=1.25in,clip,keepaspectratio]{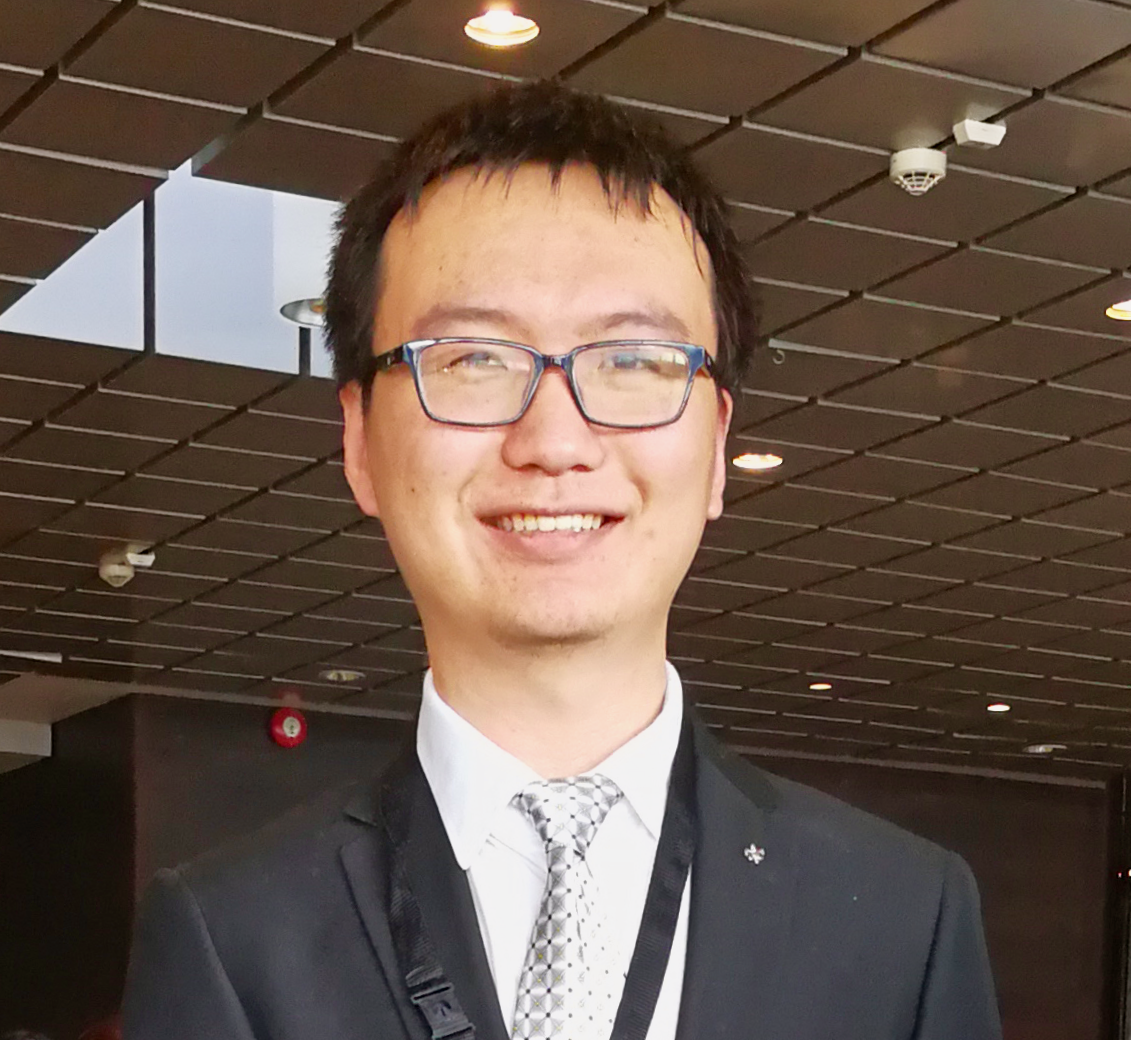}}]{Yuanhao Cui}
	received his B.S. degree in Henan University in 2012. From 2016 until 2018, he was working as a Ph.D visitor of Aalto University in Helsinki, Finland. He is currently working toward the Ph.D. degree in signal processing at  Beijing University of Posts and Telecommunications. His research interests include blockchain application in Internet of Things, coexistence of radar and communication and machine learning application in wireless communication.
\end{IEEEbiography}

\begin{IEEEbiography}[{\includegraphics[width=1in,height=1.25in,clip,keepaspectratio]{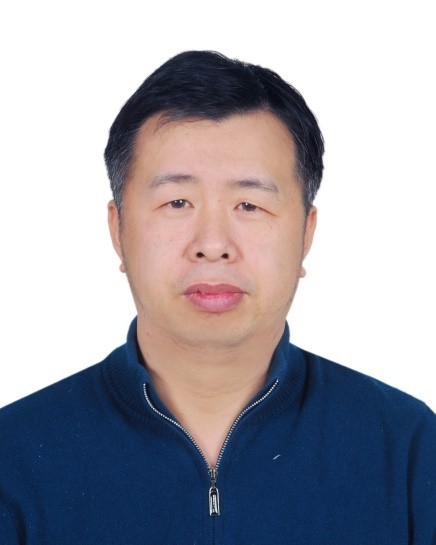}}]{Xuesong Qiu}
	received the Ph.D. degree from the Beijing University of Posts and Telecommunications, Beijing, China, in 2000. He is currently a Professor, a doctoral supervisor, and the deputy director of the State Key Laboratory of networking and switching technology, and vice president of Network Technology Research Institute of Beijing University of Posts and Telecommunications, China. He was editor of the two ITU-T standards and 4 industry standards of China. He has published more than 200 academic papers. He won twice the national scientific and technological progress prize of china. His major research interests include network and service management, information and communication technology of smart grid. Email: xsqiu@bupt.edu.cn.
\end{IEEEbiography}

\begin{IEEEbiography}[{\includegraphics[width=1in,height=1.25in,clip,keepaspectratio]{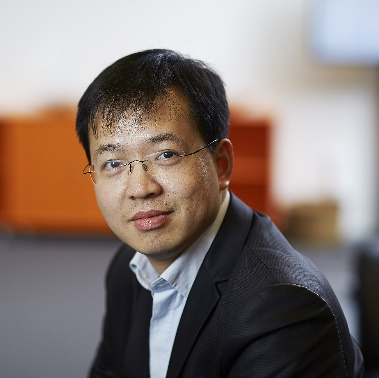}}]{Yan Zhang}(Fellow, IEEE) received the Ph.D. degree from the School of Electrical and Electronics Engineering, Nanyang Technological University, Singapore. He is currently a Full Professor with the Department of Informatics, University of Oslo, Oslo, Norway. His research interests include next-generation wireless networks leading to 5 G be-yond/6 G, green and secure cyber-physical systems (e.g., smart grid and transport). Dr. Zhang is a member of CCF Technical Committee of Blockchain. He is the Chair of IEEE Communications Society Technical Committee on Green Communications and Computing (TCGCC). He is an Editor for IEEE publications, including IEEE Communications Magazine, IEEE Network Magazine, IEEE TRANSACTIONS ON VEHICULAR TECHNOLOGY, IEEE TRANSACTIONS ON INDUSTRIAL INFORMATICS, IEEE TRANSACTIONS ON GREEN COMMUNICATIONS AND NETWORKING, IEEE COMMUNICATIONS SURVEY AND TUTORIALS, IEEE INTERNET OF THINGS JOURNAL, IEEE SYSTEMS JOURNAL, IEEE VEHICULAR TECHNOLOGY MAGAZINE, and IEEE BLOCKCHAIN TECHNI-CAL BRIEFS. He is a Symposium/Track Chair for a number of conferences, including IEEE International Conference on Communications 2021, IEEE Global Communications Conference 2017, IEEE International Symposium on Personal, Indoor and Mobile Radio Communications 2016, IEEE International Conference on Communications, Control, and Computing Technologies for Smart Grids 2015. He is an IEEE Vehicular Technology Society Distinguished Lecturer during 2016-2020 and he is named as CCF 2019 Distinguished Speaker. Since 2018, he was the recipient of the Highly Cited Researcher Award (Web of Science top 1\% most cited) by Clarivate Analytics.
\end{IEEEbiography}

%\bibliographystyle{IEEEtranTIE}
%\bibliography{ref}

%Formatter's Note: If you have any biographies to make, please include them here.
%\begin{IEEEbiography}{Michael Shell}
%Biography text here.
%\end{IEEEbiography}

\end{document}